\documentclass[aps,prd,preprint,superscriptaddress,showpacs,floatfix,11pt]{revtex4}

\usepackage[dvips]{graphics}
\usepackage[dvips]{graphicx}
\usepackage{psfrag}
\usepackage{longtable}
\usepackage{amsfonts}
\usepackage{mathrsfs}

\begin{document}

\preprint{BNL-HET-04/19, COLO-HEP-505}
\preprint{September, 2004}
\vskip 1.5in

\title{
Relating Leptogenesis to Low Energy Flavor Violating Observables in Models with Spontaneous CP Violation}


\author{Mu-Chun Chen}
\email[]{chen@quark.phy.bnl.gov}
\affiliation{High Energy Theory Group, Department of Physics, 
Brookhaven National Laboratory, Upton, NY 11973, U.S.A.}
\author{K.T. Mahanthappa}
\email[]{ktm@pizero.colorado.edu}


\affiliation{Department of Physics, University of Colorado, 
Boulder, CO80309-0390, U.S.A.}




\begin{abstract}

In the minimal left-right symmetric model, 
there are only two intrinsic CP violating phases to account 
for all CP violation in both the quark and lepton sectors, 
if CP is broken spontaneously by the complex phases 
in the VEV's of the scalar fields. In addition, the left- 
and right-handed Majorana mass terms for the neutrinos 
are proportional to each other due to the parity in the model. 
This is thus a very constrained 
framework, making the existence of correlations among 
the CP violation in leptogenesis, neutrino oscillation 
and neutrinoless double beta decay possible. In these models, 
CP violation in the leptonic sector and CP violation in the quark 
sector are also related. 
We find, however, that such connection is rather 
weak due to the large hierarchy in the bi-doublet VEV required 
by a realistic quark sector.

\end{abstract}

\maketitle

\section{Introduction}

The evidence of non-zero neutrino masses opens up the possibility that the leptonic CP violation might be responsible, through leptogenesis, for the observed asymmetry between matter and anti-matter in the 
Universe~\cite{Fukugita:1986hr}. 
It is generally difficult, however, to make connection between leptogenesis and CP-violating processes at low energies due to the presence of extra phases and mixing angles in the right-handed neutrino sector as in 
models~\cite{Joshipura:2001ui}. 
(For realistic neutrino mass models based on SUSY GUTs see, 
for example, \cite{Chen:2000fp}.) 
Recently attempts have been made to induce {\it spontaneous CP violation}  (SCPV)  from a single source. In one such attempt SM is extended by a singlet scalar field which develops a complex VEV which breaks CP 
symmetry~\cite{Branco:2003rt}. 
Another attempt assumes that there is one complex VEV of the field which breaks the $B - L$ symmetry in SO(10)~\cite{Achiman:2004qf}. 
In these models there is no compelling reason why all other VEVs have to be real. Here we focus on the minimal left-right symmetric model. In this model SCPV could be due to two intrinsic CP violating phases associated with VEVs of two scalar fields which account for all CP-violating processes observed in Nature; these {\it exhaust} sources of CP-violation. 
As the left-handed (LH) and right-handed (RH)
Majorana mass matrices are identical up to an overall mass scale, in this model there exist relations between low energy processes, such as neutrino oscillations, neutrinoless double beta decay and lepton flavor violating charged lepton decay, and leptogenesis which occurs at very high energy. Also, it is possible to relate CP-violation in the lepton sector with that in the quark sector. In this paper we explicitly display such relations in two realistic models.

The paper is organized as follows: In Sec.~\ref{model}, we define the minimal 
left-right symmetric model and review  
the formulation of leptogenesis and low energy LFV processes, 
including CP violation in neutrino oscillation and 
neutrinoless double beta decay; we then propose in Sec.~\ref{nu} 
a new model which gives rise to bi-large leptonic mixing 
patterns due to an interplay of both type-I and type-II 
see-saw terms; we also extract the connections between 
various LFV processes in this model and a flavor ansatz proposed earlier; 
Sec.~\ref{cond} concludes this paper.

\section{The Minimal Left-Right Symmetric Model}\label{model}

The minimal left-right symmetric $SU(2)_{L} \times SU(2)_{R} \times 
U(1)_{B-L}$ model~\cite{Pati:1974yy,Mohapatra:1986uf}  
is the minimal extension of the SM. 
It has the following particle content: The left- and right-handed 
matter fields transform as doublets of $SU(2)_{L}$ and $SU(2)_{R}$, 
respectively,
\begin{eqnarray}
Q_{i,L} & = \left(
\begin{array}{c}
u \\
d
\end{array}
\right)_{i,L}
\sim (1/2, 0, 1/3), \qquad 
Q_{i,R}
& = 
\left(
\begin{array}{c}
u \\
d\end{array}
\right)_{i,R}
\sim(0, 1/2, 1/3)
\\
L_{i,L}
& =   
\left(
\begin{array}{c}
e \\
\nu\end{array}
\right)_{i,L}
\sim (1/2, 0, -1)
, \qquad 
L_{i,R}
& = 
\left(
\begin{array}{c}
e \\
\nu\end{array}
\right)_{i,R}
\sim (0, 1/2, -1) \; .
\end{eqnarray}
The minimal Higgs sector that breaks the left-right symmetry 
to the SM gauge group contains a $SU(2)$ bi-doublet Higgs 
and two $SU(2)$ triplet Higgses, 
\begin{eqnarray}
\Phi & = &
\left(
\begin{array}{cc}
\phi_{1}^{0} & \phi_{2}^{+}\\
\phi_{1}^{-} & \phi_{2}^{0}\end{array}
\right)
\sim (1/2, 1/2, 0) \\
\Delta_{L} & = &
\left(
\begin{array}{cc}
\Delta^{+}_{L}/\sqrt{2} & \Delta^{++}_{L} \\
\Delta^{0}_{L} & -\Delta^{+}_{L}/\sqrt{2}
\end{array}\right)
\sim (1, 0, +2) \\
\Delta_{R} & = &
\left(
\begin{array}{cc}
\Delta^{+}_{R}/\sqrt{2} & \Delta^{++}_{R} \\
\Delta^{0}_{R} & -\Delta^{+}_{R}/\sqrt{2}
\end{array}\right)
\sim (0,1,+2) \; .
\end{eqnarray}
The $SU(2)_{R}$ symmetry is broken by the VEV of the triplet $\Delta_{R}$,
\begin{equation}
<\Delta_{R}> = 
\left(
\begin{array}{cc}
0 & 0 \\
v_{R} e^{i\alpha_{R}} & 0
\end{array}\right) \; .
\end{equation}
and the electroweak symmetry is broken by the VEV of the bi-doublet,
\begin{equation}
<\Phi> = 
\left(
\begin{array}{cc}
\kappa e^{i\alpha_{\kappa}} & 0\\
0 & \kappa^{'}e^{i\alpha_{\kappa^{'}}}
\end{array}
\right) \; .
\end{equation}
To get realistic SM gauge boson masses, the VEV's of the bi-doublet Higgs 
must satisfy 
$v^{2} \equiv 
|\kappa|^2 + |\kappa^{'}|^{2} \simeq 2 m_{w}^{2}/g^{2} \simeq (174 GeV)^{2}$.
Generally, a non-vanishing VEV for the 
$SU(2)_{L}$ triplet Higgs is induced, and it is suppressed 
by the heavy $SU(2)_{R}$ breaking scale similar 
to the see-saw mechanism for the neutrinos,  
\begin{equation}
<\Delta_{L}> =  \left(
\begin{array}{cc}
0 & 0 \\
v_{L} e^{i\alpha_{L}} & 0
\end{array}\right) \; , \qquad v_{L}v_{R} 
= \beta |\kappa|^{2}  \; ,
\end{equation}
where the parameter $\beta$ is a function of the order 
$\mathcal{O}(1)$ coupling constants in the scalar potential 
and $v_{R}$, $v_{L}$, $\kappa$ and $\kappa^{\prime}$ 
are positive real numbers in the above equations. 
Due to this see-saw suppression, for a $SU(2)_{R}$ breaking scale as high as 
$10^{15}\; GeV$ which is required by the smallness of the neutrino masses,  
the induced $SU(2)_{L}$ triplet VEV is well below the upper bound 
set by the electroweak precision 
constraints~\cite{Chen:2003fm}.  
The scalar potential that gives rise to the 
vacuum alignment described can be found in 
Ref.~\cite{Deshpande:1990ip,Rodriguez:2002ey}.

The Yukawa sector of the model is given by
$\mathcal{L}_{Yuk} = \mathcal{L}_{q} + \mathcal{L}_{\ell}$
where $\mathcal{L}_{q}$ and $\mathcal{L}_{\ell}$ are 
the Yukawa interactions in the quark and lepton sectors, 
respectively. The Lagrangian for quark Yukawa interactions is 
given by,  
\begin{equation}
-\mathcal{L}_{q} = \overline{Q}_{i,R} (F_{ij} \Phi + G_{ij} 
\tilde{\Phi}) Q_{j,L} + h.c.
\end{equation}
where $\tilde{\Phi} \equiv \tau_{2} \Phi^{\ast} \tau_{2}$. 
In general, $F_{ij}$ and $G_{ij}$ 
are Hermitian to preseve left-right symmetry. Because of our assumption  of 
SCPV with complex vacuum expectation values, 
the matrices $F_{ij}$ and $G_{ij}$ are real. 
The Yukawa interactions responsible for generating the lepton masses 
are summarized in the following  Lagrangian, 
$\mathcal{L}_{\ell}$,
\begin{equation}
-\mathcal{L}_{\ell} =  
\overline{L}_{i,R} (P_{ij} \Phi + R_{ij} \tilde{\Phi}) L_{j,L} 
+ i f_{ij} (L_{i,L}^{T} \mathcal{C}\tau_{2} \Delta_{L} L_{j,L} 
+ L_{i,R}^{T} C\tau_{2} \Delta_{R} L_{j,R}) 
+ h.c. \; ,
\end{equation}
where $\mathcal{C}$ is the Dirac charge conjugation operator, and the matrices 
$P_{ij}$, $R_{ij}$ and $f_{ij}$ are real due to the assumption of SCPV.
Note that the Majorana mass terms $L_{i,L}^{T} \Delta_{L} L_{j,L}$ and 
$L_{i,R}^{T} \Delta_{R} L_{j,R}$ have identical coupling because 
the Lagrangian must be invariant under interchanging 
$L \leftrightarrow R$.
The complete Lagrangian of the model is invariant under 
the unitary transformation, 
under which the matter fields transform as
\begin{equation}
\psi_{L} \rightarrow U_{L} \psi_{L}, \qquad
\psi_{R} \rightarrow U_{R} \psi_{R}
\end{equation}
where $\psi_{L,R}$ are left-handed (right-handed) fermions, 
and the scalar fields transform according to
\begin{equation}
\Phi \rightarrow U_{R} \Phi U_{L}^{\dagger}
, \qquad 
\Delta_{L} \rightarrow U_{L}^{\ast} \Delta_{L} U_{L}^{\dagger}
, \qquad 
\Delta_{R} \rightarrow U_{R}^{\ast} \Delta_{R} U_{R}^{\dagger}
\end{equation}
with the unitary transformations $U_{L}$ and $U_{R}$ being 
\begin{equation}\label{unit}
U_{L}  =  
\left(
\begin{array}{cc}
e^{i\gamma_{L}} & 0\\
0 & e^{-i\gamma_{L}}
\end{array}
\right)
, \qquad 
U_{R} = 
\left(
\begin{array}{cc}
e^{i\gamma_{R}} & 0\\
0 & e^{-i\gamma_{R}}
\end{array}
\right) \; .
\end{equation}
Under these unitary transformations, the VEV's transform as
\begin{equation}
\kappa  \rightarrow  \kappa e^{-i(\gamma_{L}-\gamma_{R})}
, \quad 
\kappa^{\prime} \rightarrow  \kappa^{\prime} e^{i(\gamma_{L}-\gamma_{R})}
, \quad 
v_{L}  \rightarrow  v_{L} e^{-2i\gamma_{L}}
, \quad 
v_{R} \rightarrow  v_{R} e^{-2i\gamma_{R}} \; .
\end{equation}
Thus by re-defining the phases of matter fields with the choice of  
$\gamma_{R}  =  \alpha_{R}/2$ and
$\gamma_{L} =   \alpha_{\kappa} + \alpha_{R}/2$ 
in the unitary matrices $U_{L}$ and $U_{R}$, 
we can rotate away two of the complex phases in the VEV's of 
the scalar fields and are left with only two genuine CP violating phases, 
$\alpha_{\kappa^\prime}$ and $\alpha_{L}$, 
\begin{equation}
<\Phi>  = \left(
\begin{array}{cc}
\kappa & 0\\
0 & \kappa^{\prime}e^{i\alpha_{\kappa^{\prime}}}
\end{array}
\right), \quad
<\Delta_{L}>  =  
\left(
\begin{array}{cc}
0 & 0 \\
v_{L}e^{i\alpha_{L}} & 0
\end{array}\right), \quad
<\Delta_{R}> = 
\left(
\begin{array}{cc}
0 & 0 \\
v_{R} & 0
\end{array}\right).
\end{equation}

The quark Yukawa interaction $\mathcal{L}_{q}$ gives rise 
to quark masses after the bi-doublet acquires VEV's
\begin{equation}
M_{u} = \kappa F_{ij} + \kappa^{\prime}  
e^{-i \alpha_{\kappa^\prime}} G_{ij}, 
\quad 
M_{d} = \kappa^{\prime} e^{i\alpha_{\kappa^\prime}}  F_{ij} 
+ \kappa G_{ij} \; .
\end{equation}
Thus the relative phase in the two VEV's in the SU(2) 
bi-doublet, $\alpha_{\kappa^\prime}$, gives rise 
to the CP violating phase in the CKM matrix. 
To obtain realistic quark masses and CKM matrix elements, 
it has been shown that the VEV's of the bi-doublet 
have to satisfy $\kappa/\kappa^\prime \simeq m_{t}/m_{b} \gg 1$~\cite{Ball:1999mb}.
When the triplets and the bi-doublet acquire VEV's, we obtain the following 
mass terms for the leptons
\begin{eqnarray}
M_{e} = \kappa^{\prime} e^{i\alpha_{\kappa^\prime}} P_{ij} + \kappa R_{ij}, 
& \quad
M_{\nu}^{Dirac} = \kappa P_{ij} 
+ \kappa^{\prime} e^{-i\alpha_{\kappa^\prime}} R_{ij} \\
M_{\nu}^{RR} = v_{R} f_{ij}, & \quad
M_{\nu}^{LL} = v_{L} e^{i\alpha_{L}} f_{ij} \; .
\end{eqnarray}
The effective neutrino mass matrix, $M_{\nu}^{\mbox{\tiny eff}}$, which 
arises from the Type-II seesaw mechanism~\cite{Mohapatra:1979ia},  
is thus given by
\begin{eqnarray}
M_{\nu}^{\mbox{\tiny eff}} & = & M_{\nu}^{II} - M_{\nu}^{I}\\
M_{\nu}^{I} & = &  
(M_{\nu}^{\mbox{\tiny Dirac}})^{T} (M_{\nu}^{RR})^{-1} 
(M_{\nu}^{\mbox{\tiny Dirac}})
\nonumber\\
& = &
 (\kappa P + \kappa^{\prime} e^{-i \alpha_{\kappa^{\prime}}} R)^{T}
(v_{R} f)^{-1}
(\kappa P + \kappa^{\prime} e^{-i \alpha_{\kappa^{\prime}}} R)  
\\
M_{\nu}^{II} & = & v_{L} e^{i\alpha_{L}} f \; .
\end{eqnarray}

Assuming the charged lepton mass matrix is diagonal, 
the Yukawa couplings $R_{ij}$ can be determined  
by the charged lepton masses, 
\begin{equation}
R = \left(\begin{array}{ccc}
\mathcal{O}(m_{e}/m_{\tau}) & 0 & 0\\
0 &  \mathcal{O}(m_{\mu}/m_{\tau}) & 0\\
0 & 0 & \mathcal{O}(1)
\end{array}\right) \; .
\end{equation}
In the limit $\kappa \gg \kappa^\prime$, the 
conventional type-I see-saw term~\cite{type-I,Mohapatra:1979ia} 
is dominated by the term proportional to $\kappa$,
\begin{equation}
M_{\nu}^{I} = (M_{\nu}^{\mbox{\tiny Dirac}})^{T} 
(M_{\nu}^{RR})^{-1} (M_{\nu}^{\mbox{\tiny Dirac}})
\simeq 
\frac{\kappa^{2}}{v_{R}}P^{T} f^{-1} P 
= \frac{v_{L}}{\beta}P^{T} f^{-1} P
\; .
\end{equation}
Consequently, the connection between CP violation in the quark 
sector and that in the lepton sector, which is 
made through the phase $\alpha_{\kappa^{\prime}}$, appears only 
at the sub-leading order, $\mathcal{O}
\left( \kappa^{\prime}/\kappa \right)$, thus making this connection 
rather weak. It has also been shown in Ref.~\cite{Rodriguez:2002ey} 
that in order to avoid flavor changing neutral current, 
the phase $\alpha_{\kappa^{\prime}}$ 
has to be close to zero. In this case, leptonic CP violation is not
constrained by $\alpha_{\kappa^{\prime}}$, and thus it can be
large (due to non-zero $\alpha_{L}$). 
We will neglect these sub-leading order terms in this paper.  
In this case there is thus only one phase, 
$\alpha_{L}$, that is responsible for all leptonic CP 
violation. 
As the charged lepton mass matrix is diagonal, 
the leptonic mixing matrix, the so-called Maki-Nakagawa-Sakata 
(MNS) matrix,  is obtained by diagonalizing the effective 
neutrino mass matrix
$M_{\nu}^{\mbox{\tiny diag}} 
= U_{\mbox{\tiny MNS}}^{\dagger} M_{\mbox{\tiny eff}}^{\nu} 
U_{\mbox{\tiny MNS}}^{\ast}
= \mbox{diag} \left( m_{\nu_{1}}, m_{\nu_{2}}, m_{\nu_{3}} \right)$,
where $m_{\nu_{1,2,3}}$ are real and positive, 
and it can be parameterized as the product of a CKM-like mixing matrix, 
which has three mixing angles and one CP violating phase, with 
a diagonal phase matrix,  
\begin{eqnarray}
U_{MNS} = 
\left(
\begin{array}{ccc}
c_{12}c_{13} &
s_{12}c_{13} &
s_{13}\\
-s_{12}c_{23}-c_{12}s_{23}s_{13}e^{i\delta_{\ell}} &
c_{12}c_{23}-s_{12}s_{23}s_{13}e^{i\delta_{\ell}} &
s_{23}c_{13}e^{i\delta_{\ell}}\\
s_{12}s_{23}-c_{12}c_{23}s_{13}e^{i\delta_{\ell}} &
-c_{12}s_{23}-s_{12}c_{23}s_{13}e^{i\delta_{\ell}} &
c_{23}c_{13}e^{i\delta_{\ell}}
\end{array}
\right) 
{\scriptstyle \times} \left(
\begin{array}{ccc}
1 & &\\
& e^{i \frac{\alpha_{21}}{2}} & \\
& & e^{i\frac{\alpha_{31}}{2}}
\end{array}
\right) \; . 
\end{eqnarray}
The Dirac phase, $\delta_{\ell}$, as well as the two Majorana phases, 
$\alpha_{21}$ and $\alpha_{31}$, are given in terms a single phase, 
$\alpha_{L}$, in this model. The relations among these three leptonic CP 
violating phases thus ensue. The analytic relations between $\alpha_{L}$ 
and the three leptonic CP phases are very complicated, 
and are not explicitly expressed. The effects of these three 
CP phases appear in the following processes.

\subsection{CP Violation in Neutrino Oscillation} 

The Dirac CP violating phase, $\delta_{\ell}$, affects  
neutrino oscillation. The transition probability of the flavor eigenstate 
$\nu_{\alpha}$ into $\nu_{\beta}$ 
( $\alpha, \; \beta = e, \; \mu, \; \tau$) reads, 
\begin{equation}
P(\nu_{\alpha} \rightarrow \nu_{\beta}) 
= \delta_{\alpha\beta} 
- 4 \sum_{i > j} Re (U_{\alpha i} U_{\beta j} U_{\alpha j}^{\ast} 
U_{\beta i}^{\ast} ) \sin^{2} (\Delta m_{ij}^{2} \frac{L}{4E}) 
+ 2 \sum_{i > j} J_{\mbox{\tiny 
CP}} \sin^{2} (\Delta m_{ij}^{2} \frac{L}{4E})
\end{equation}
where the CP violation is governed by the leptonic Jarlskog invariant, 
$J_{\mbox{\tiny CP}}$, 
which can be expressed model-independently 
in terms of the effective neutrino mass matrices as~\cite{Branco:2002xf}
\begin{equation}
J_{\mbox{\tiny CP}} = 
-\frac{\mbox{Im}(H_{12}H_{23}H_{31})}{\Delta m_{21}^{2} 
\Delta m_{31}^{2} \Delta m_{32}^{2}} \;, \qquad 
H \equiv M_{\nu}^{\mbox{\tiny eff}} M_{\nu}^{\mbox{\tiny eff}\dagger} \; ,
\end{equation}
where $\Delta m_{ij}^{2} \equiv m_{i}^{2} - m_{j}^{2}$ 
$(i,j=1,2,3)$ with $m_{i}$ being the 
mass eigenvalues of the effective neutrino mass matrix, 
$M_{\nu}^{\mbox{\tiny eff}}$. 

\subsection{Neutrinoless Double Beta Decays}

Neutrinoless double beta ($0\nu\beta\beta$) 
decay is, on the other hand, only sensitive to 
the two Majorana phases, $\alpha_{21}$ and $\alpha_{31}$. 
Their dependence in the $0\nu\beta\beta$ matrix element, $\left< 
m_{ee} \right>$, is
\begin{eqnarray}
\left| \left< m_{ee} \right> \right|^{2} & = & 
m_{1}^{2} \left| U_{e1} \right|^{4} +   m_{2}^{2} \left| U_{e2} \right|^{4}  
+ m_{3}^{2} \left| U_{e3} \right|^{4}
+ 2 m_{1} m_{2} \left| U_{e1} \right|^{2} 
\left| U_{e2} \right|^{2} \cos \alpha_{21}
\nonumber\\
 & & + 2 m_{1} m_{3} \left| U_{e1} \right|^{2} 
\left| U_{e3} \right|^{2} \cos \alpha_{31}
+ 2 m_{2} m_{3} \left| U_{e2} \right|^{2} \left| U_{e3} 
\right|^{2} \cos \left( \alpha_{31} - \alpha_{21} \right) \; ,
\end{eqnarray}
where $U_{e i}$ ($i = 1,2,3$) are the matrix elements in the first row 
of the MNS matrix.

\subsection{Leptogenesis}
 
In the left-right symmetric model with the particle content we 
have, leptogenesis receives contributions both from 
the decay of the lightest RH neutrino, $N_{1}$, as well as 
from the decay of the $SU(2)_{L}$ triplet Higgs, 
$\Delta_{L}$~\cite{Joshipura:2001ya,Antusch:2004xy}. 
We consider the $SU(2)_{L}$ triplet Higgs 
being heavier than the lightest RH neutrino, $M_{\Delta_{L}} > M_{R_{1}}$. 
For this case, the decay of the lightest RH neutrino dominates.
In the SM, the canonical contribution to the lepton number 
asymmetry from one-loop diagrams 
mediated by the Higgs doublet and the charged leptons 
is given by~\cite{Antusch:2004xy},    
\begin{equation}
\epsilon^{N_{1}}  =  
\frac{3}{16\pi }  \biggl( \frac{M_{R_{1}}  }{v^{2}} \biggr)  
\cdot \frac{
\mbox{Im} \biggl(  \mathcal{M}_{D}  \left( M_{\nu}^{I} \right)^{\ast} 
\mathcal{M}_{D}^{T} 
\biggr)_{11} }{ ( \mathcal{M}_{D} \mathcal{M}_{D}^{\dagger} )_{11} }  \; . 
\end{equation}
Now, there is one additional one-loop diagram mediated 
by the $SU(2)_{L}$ triplet Higgs.  It contributes to the 
decay amplitude of the right-handed neutrino into a doublet 
Higgs and a charged lepton, which 
gives an additional contribution to the lepton number 
asymmetry~\cite{Antusch:2004xy},   
 \begin{equation}
 \epsilon^{\Delta_{L}}  =  
\frac{3}{16\pi }  \biggl( \frac{M_{R_{1}}  }{v^{2}} \biggr)  
\cdot \frac{
 \mbox{Im} \biggl(  \mathcal{M}_{D}  \left( M_{\nu}^{II} \right)^{\ast} 
\mathcal{M}_{D}^{T} 
 \biggr)_{11} }{ ( \mathcal{M}_{D} \mathcal{M}_{D}^{\dagger} )_{11} }  \; ,
 \end{equation} 
where $\mathcal{M}_{D}$ is the neutrino Dirac mass term in the basis where 
the RH neutrino Majorana mass term is real and diagonal, 
\begin{equation}
\mathcal{M}_{D} = O_{R} M_{D}, \; \quad 
f^{\mbox{\tiny diag}} = O_{R} f O_{R}^{T} \; .
\end{equation}  
Because there is no phase present in either 
$M_{D} = P \kappa$ or $M_{\nu}^{I}$ or $O_{R}$, 
the quantity $ \mathcal{M}_{D} \left( M_{\nu}^{I} \right)^{\ast} 
\mathcal{M}_{D}^{T}$ is real, leading to a vanishing $\epsilon^{N_{1}}$. 
We have checked explicitly that this statement is 
true for {\it any} chosen unitary transformations $U_{L}$ and $U_{R}$ 
defined in Eq.~(\ref{unit}).  
On the other hand, the contribution, 
$\epsilon^{\Delta_{L}}$, due to the diagram mediated by 
the $SU(2)_{R}$ triplet is proportional to $\sin\alpha_{L}$. 
So, as long as the phase $\alpha_{L}$ is non-zero, 
the predicted value for $\epsilon^{\Delta_{L}}$ is finite.   
A non-vanishing value for 
$\epsilon^{N_{1}}$ is generated at the sub-leading order when 
terms of order $\mathcal{O}(\kappa^{\prime}/\kappa)$ in $M_{D}$ are included. 
At the leading order, leptogenesis is generated solely from the decay 
mediated by the $SU(2)_{L}$ triplet Higgs.

\section{Specific Models with Bi-Large Neutrino Mixing}\label{nu}

In this section we consider two models which give bi-large neutrino mixing. (i) Model I assumes hierarchial mass ordering, $m_{3} \gg  m_{1,2}$, in the neutrino sector. Unlike most previous models in which either type-I or type-II see-saw mass term is supposed to dominate over the other, the bi-large mixing pattern arises in Model I due to an interplay between the type-I and type-II see-saw mass terms. (ii) In Model II we incorporate the flavor ansatz proposed in Ref.~\cite{Rodejohann:2004cg} into the LR model defined in Sec. II with the assumption of SCPV. In Ref.~\cite{Rodejohann:2004cg}, as the coupling constants are complex, the total number of independent phases is 12.
Now, as there is only one phase in our leptonic sector, all these 12 phases either vanish or are related leading to very pronounced correlation among CP violating processes. The main difference between these two models is their predictions for the atmospheric mixing angle; the deviation from maximal mixing is negligibly small in Model I whereas Model II has a sizable deviation.  In both models we assume the neutrino Yukawa coupling $P_{ij}$ to be proportional to the up quark mass matrix,
\begin{equation}
P_{ij} = q \left(
\begin{array}{ccc}
\frac{m_{u}}{m_{t}} & 0 & 0 \\
0 & \frac{m_{c}}{m_{t}} & 0 \\
0 & 0 & 1
\end{array}
\right) \; , 
\end{equation}
where the parameter $q$ is a constant of proportionality. 
A non-zero value for $U_{e3}$ is predicted in both models.

\subsection{Model I}

Assuming the matrix $f_{ij}$ have the following hierarchical elements,
\begin{equation}
f_{ij} = \left(\begin{array}{ccc}
t^{2} & t & -t\\
t & 1 & 1\\
-t & 1 & 1
\end{array}\right) \; , 
\end{equation}
with $t$ being a small and positive number. 
This mass matrix gives rise to bi-large mixing pattern 
with solar mixing angle given by  
$\tan \theta_{12}^{0} = \frac{1}{2} \biggl[ \frac{t}{\sqrt{2}} 
+ \sqrt{\frac{t^{2}}{2}+4} \;  \biggr]$, 
which is always greater than one (the light side region), and thus 
inconsistent with the presence of the matter effects 
observed experimentally.  But 
the contribution to the total effective neutrino mass 
matrix from the conventional 
type-I see-saw term, 
\begin{equation}
M_{\nu}^{I} = \frac{q^{2}}{2\beta} 
\left(\begin{array}{ccc}
0 & \frac{1}{t} \frac{m_{u}m_{c}}{m_{t}^{2}} & -\frac{1}{t} 
\frac{m_{u}}{m_{t}}\\
\frac{1}{t} \frac{m_{u}m_{c}}{m_{t}^{2}} & 0 & \frac{m_{c}}{m_{t}}\\
- \frac{1}{t} \frac{m_{u}}{m_{t}} & \frac{m_{c}}{m_{t}} & 0
\end{array}\right) v_{L} \; ,
\end{equation}
can reduce the solar mixing angle so that it is 
in the dark side region consistent with the solar experiment. 
For $t \sim \mathcal{O}(0.1)$, the (23) and (32) elements dominate. 
In the limit $m_{u}=0$, the resulting atmospheric mixing angle is maximal 
and the mixing angle $\sin\theta_{13}=0$ leading to 
a vanishing leptonic Jarlskog invariant, 
$J_{\mbox{\tiny CP}} = 0$. 
However, when $m_{u}$ is turned on, $\sin\theta_{13}$ acquires 
a non-zero value, which is suppressed by $(m_{u}/m_{t})$. 
It is therefore important to keep all three diagonal 
elements in the matrix $P_{ij}$ non-zero in our analysis. With the 
approximation, $m_{u} : m_{c} : m_{t} \simeq \epsilon^{8} 
: \epsilon^{4} : 1$, where $\epsilon = 0.22$ is the sine of the Cabibbo angle, to the leading order in $\epsilon$ 
the leptonic Jarlskog invariant is given by,
\begin{equation}
J_{\mbox{\tiny CP}} 
\simeq -\frac{2st^{2}\left(1-t^{2}\right)v_{L}^{6}}{
\Delta m_{21}^{2} \Delta m_{31}^{2} \Delta m_{32}^{2}}
\frac{m_{u}}{m_{t}} \sin\alpha_{L} \; .
\end{equation}

\begin{figure}
{\center
\includegraphics[scale=0.45,angle=270]{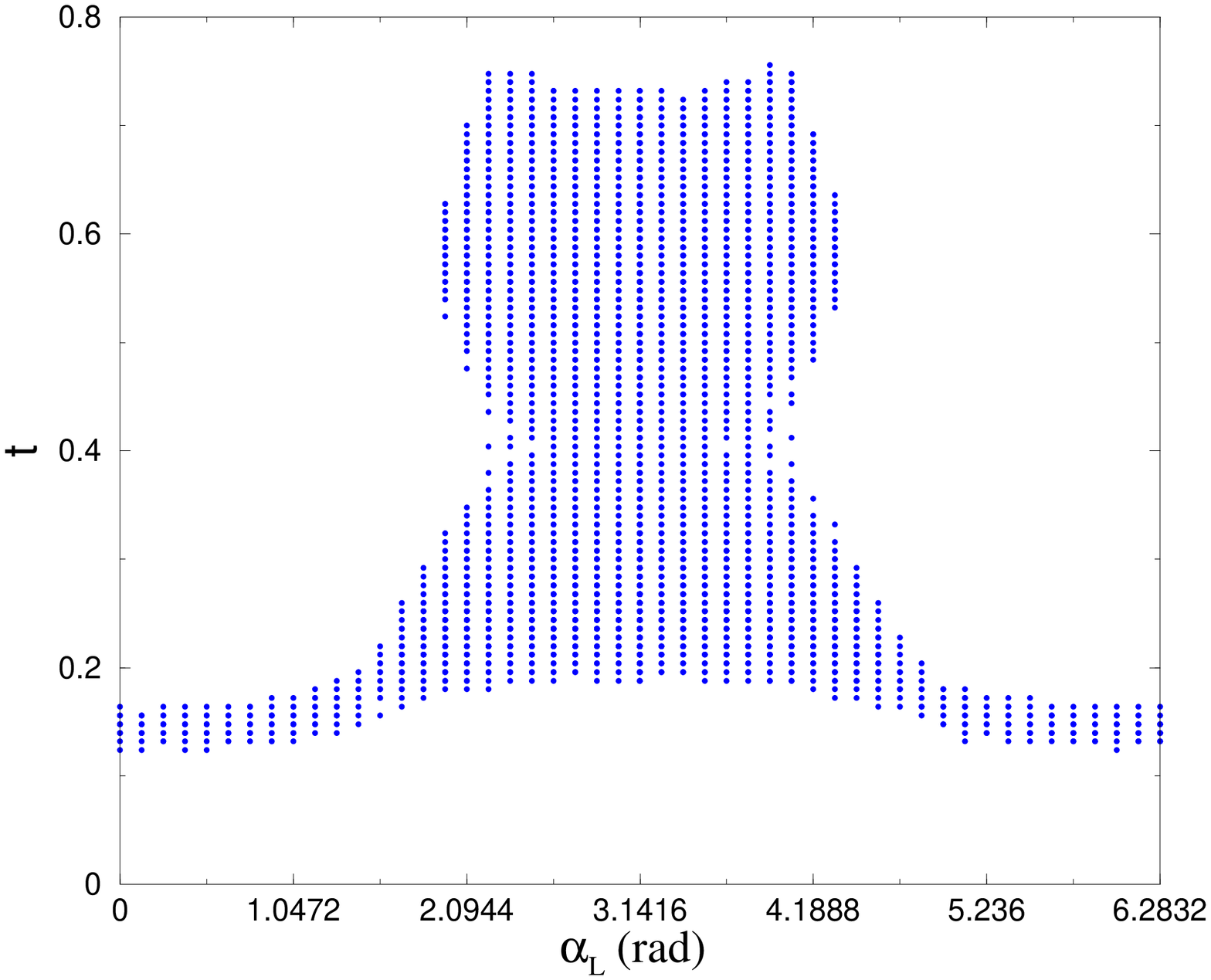}
\includegraphics[scale=0.45,angle=270]{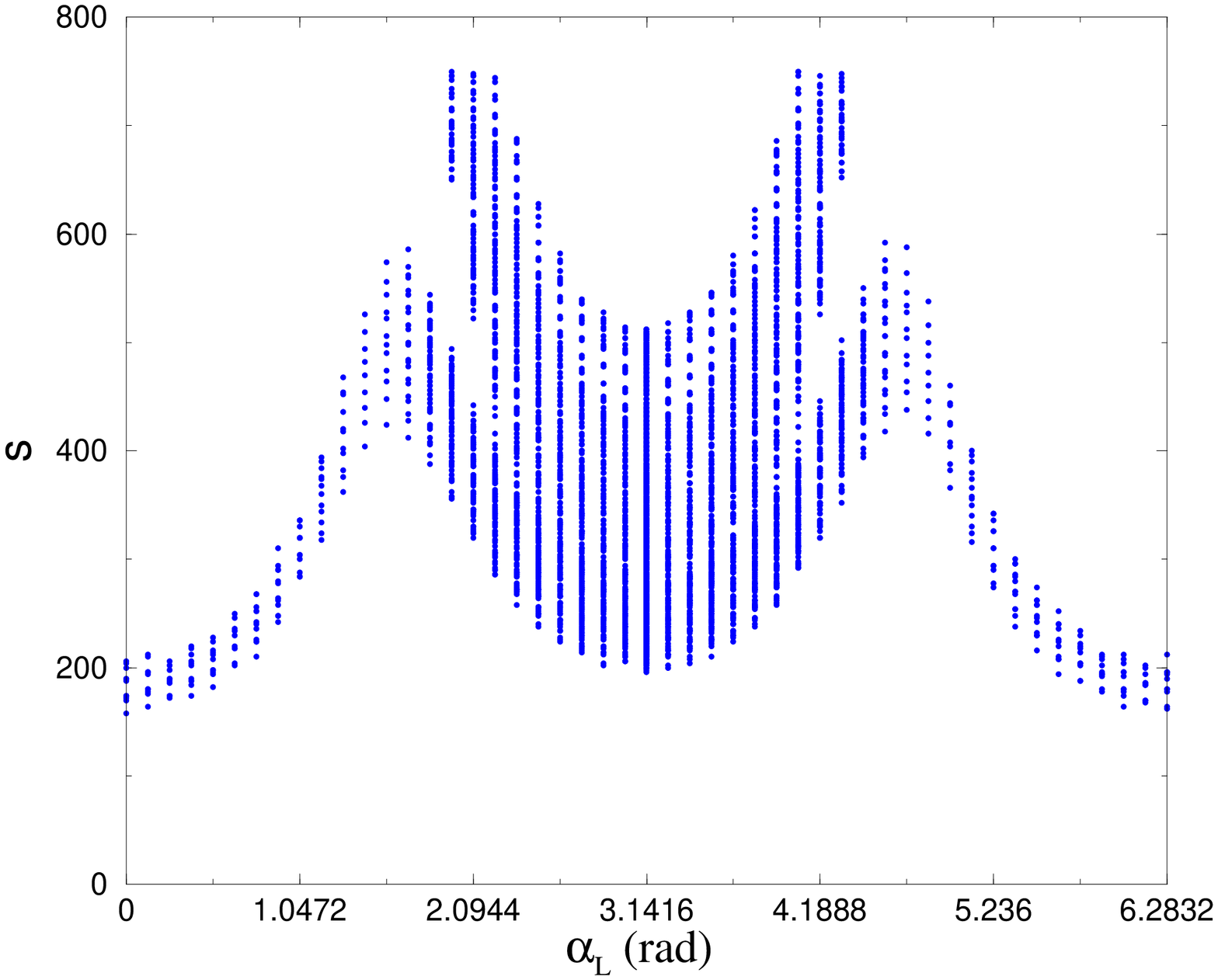}
\includegraphics[scale=0.45,angle=270]{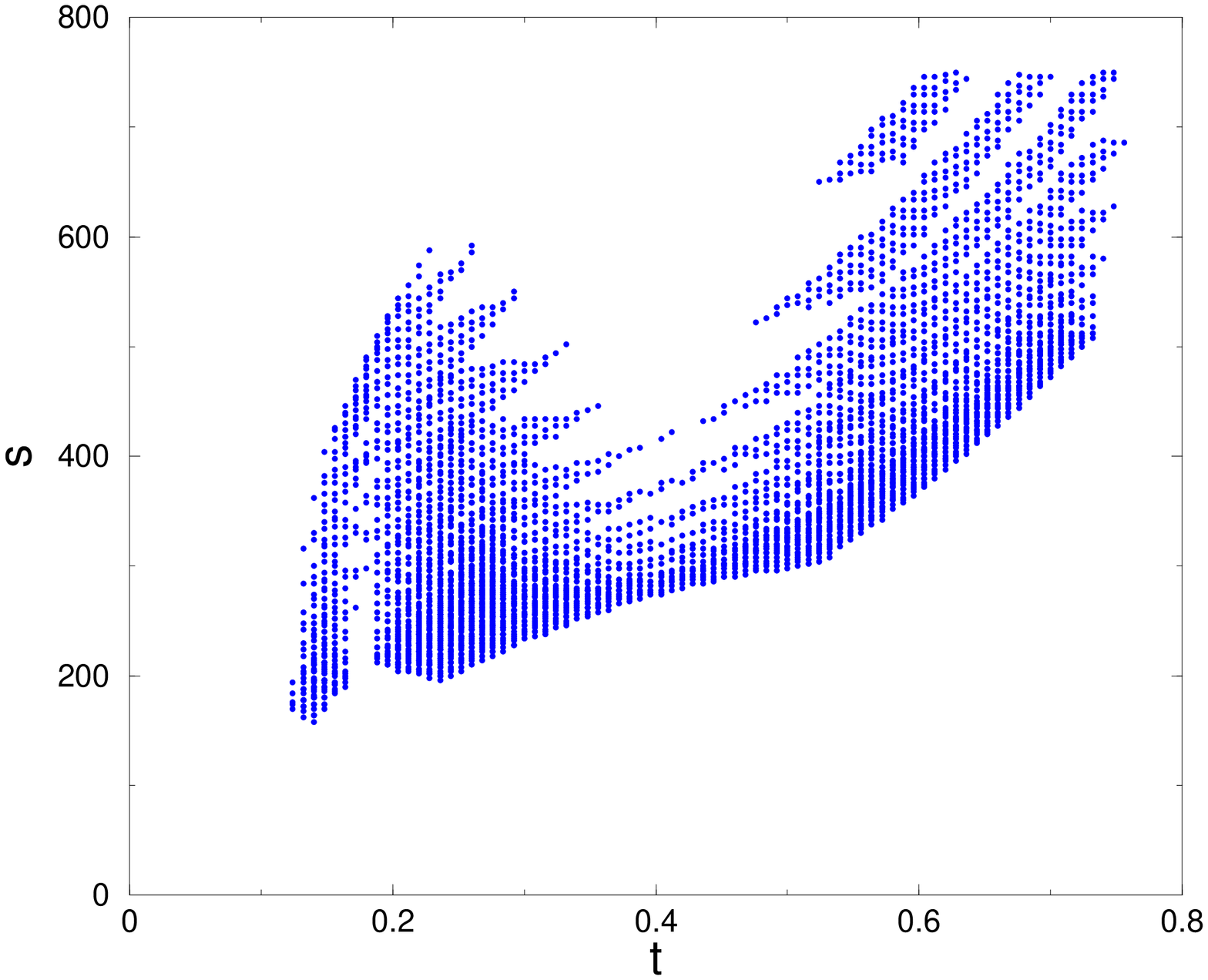}}
\caption{\label{fig-r1-a-t} Full allowed parameter space on the 
$(\alpha_{L},t,s)$ plane in Model I. 
}
\end{figure}

In our analyses, we set $m_{u}/m_{t} = (0.22)^{8}$ 
and $m_{c}/m_{t} = (0.22)^{4}$. 
As the absolute mass scale of the neutrinos does not depend on 
the parameters $(t,s,\alpha_{L})$, where  
$s$ is defined as $s=q^{2}/(2\beta)$, these parameters 
can be determined using the following neutrino oscillation 
parameters from experimental data 
at $1 \sigma$ as input~\cite{Maltoni:2004ei},
\begin{equation}\label{eq-data}
\frac{\Delta m_{\odot}^{2}}{\Delta m_{\mbox{\tiny atm}}^{2}} 
= 0.0263 \sim 0.0447 \;, \quad
\sin^{2} 2\theta_{\mbox{\tiny atm}} > 0.9 \; , \quad
\tan^{2} \theta_{\odot} = 0.35 \sim 0.44 \; .
\end{equation}
We search the {\it full} parameter space spanned by $(t,s,\alpha_{L})$, 
by allowing $\alpha_{L}$ to vary between $[0,2\pi]$, $t$ 
to vary between $[0,1]$ (so that there is normal hierarchy among the light neutrino masses), 
and $s$ to vary between $[100,1000]$ (as there are no allowed regions beyond $[100,1000]$). We slice the $(t,s,\alpha_{L})$-space given above into $(250,250,72)$ {\it equally} 
spaced points and test whether each of these points 
satisfied the constraints from the oscillation data given 
in Eq.~\ref{eq-data}. 
The allowed region for $(t,s,\alpha_{L})$ which satisfy these data 
is shown in Fig.~\ref{fig-r1-a-t}. 
The absolute scale for $v_{L}$ is not essential as it does not affect the 
qualitative behavior of the correlations, and can be changed by 
rescaling $t$ and $s$. The essential parameter that has to be taken 
into account is $r$, which differs for each data point $(t,s,\alpha_{L})$. 
We find that for all points in the allowed region given in 
Fig.~\ref{fig-r1-a-t}, by allowing  
the $SU(2)_{L}$ triplet VEV $v_{L} = r \times 
(0.0265 \; \mbox{eV})$ with $r = (0.713 - 1.16)$,  
the predicted absolute mass scales of $\Delta m_{\odot}^{2}$ 
and $\Delta m_{\mbox{\tiny atm}}^{2}$ individually 
satisfy the experimental $1\sigma$ limits, 
$\Delta m_{\mbox{\tiny atm}}^{2} = (1.9 \sim 3.0) \times 
10^{-3} \; \mbox{eV}^{2}$ and $\Delta m_{\odot}^{2} 
= (7.9 \sim 8.5) \times 10^{-5} \; 
\mbox{eV}^{2}$~\cite{Maltoni:2004ei}.

\begin{figure}
{\center
\includegraphics[scale=0.60,angle=270]{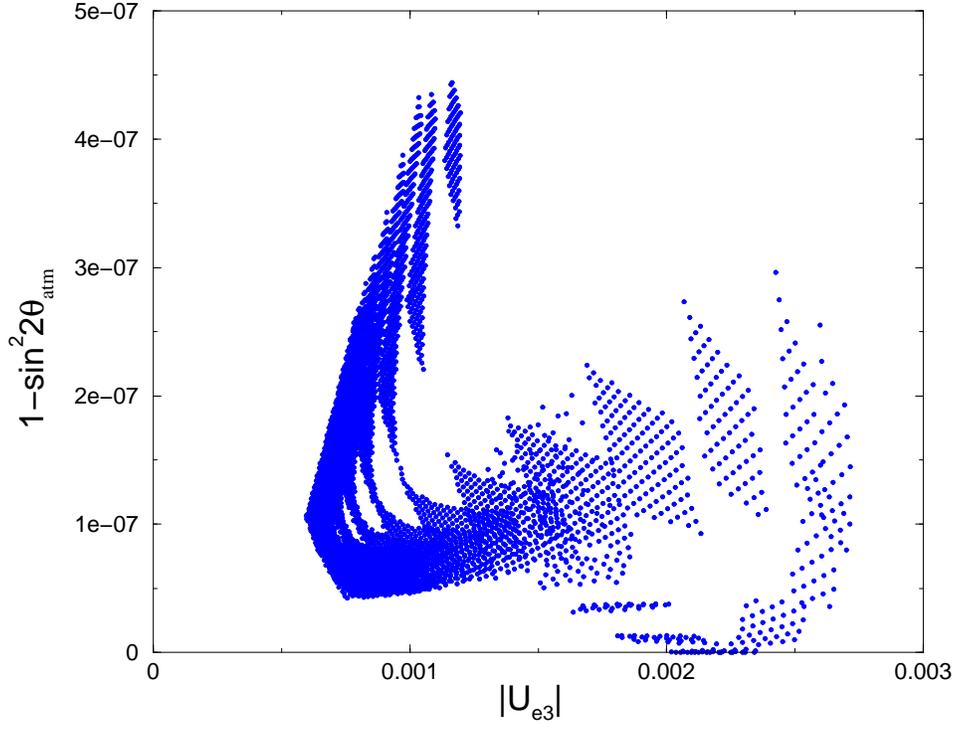}}
\caption{\label{fig-r1-ue3-uatm} The correlation between   
the leptonic mixing matrix element, $\left| 
U_{e3} \right|$, and 
the deviation of the atmospheric mixing 
angle $\sin^{2}2\theta_{\mbox{\tiny atm}}$ from $1$ in Model I.
}
\end{figure}

\begin{figure}
{\center
\includegraphics[scale=0.60,angle=270]{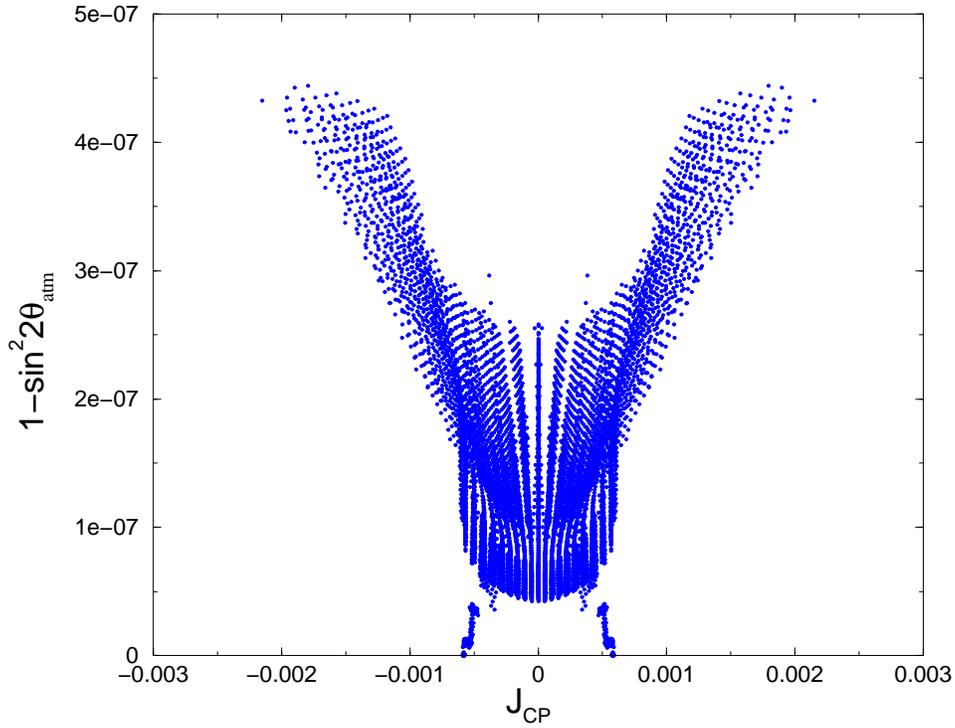}}
\caption{\label{fig-r1-Jcp-uatm} Correlation between 
the leptonic Jarlskog invariant and 
the deviation of the atmospheric mixing 
angle $\sin^{2}2\theta_{\mbox{\tiny atm}}$ from $1$ in Model I. 
}
\end{figure}

\begin{figure}
{\center
\includegraphics[scale=0.60,angle=270]{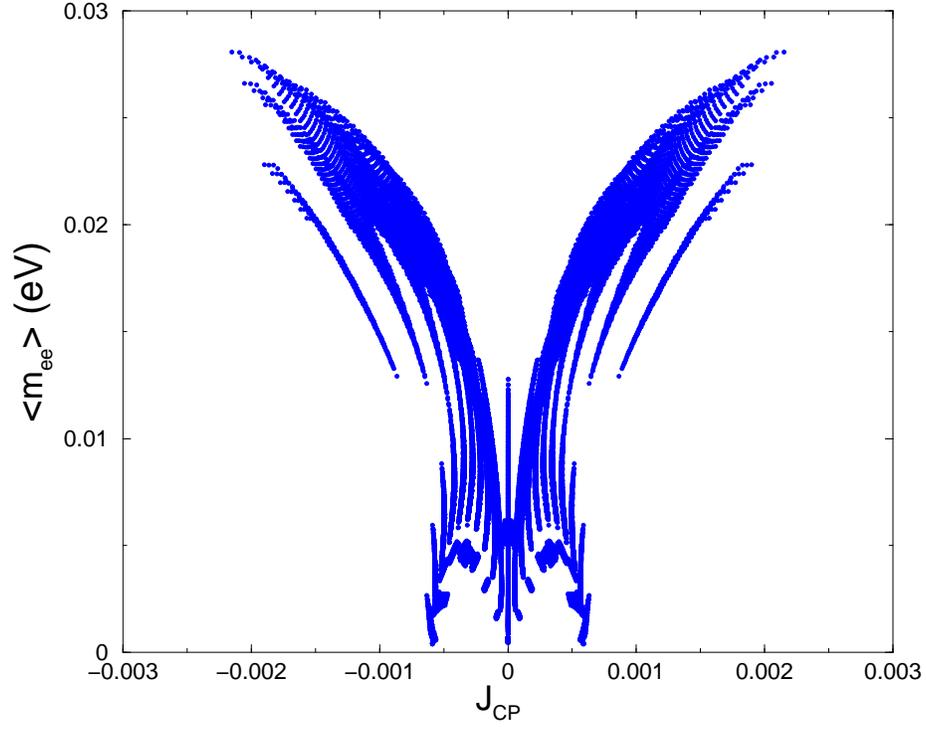}}
\caption{\label{fig-r1-Jcp-nulb} Correlation between 
the matrix element of neutrinoless double beta decay, $\left< 
m_{ee} \right>$, and the leptonic Jarlskog invariant in Model I. 
}
\end{figure}

\begin{figure}
{\center
\includegraphics[scale=0.60,angle=270]{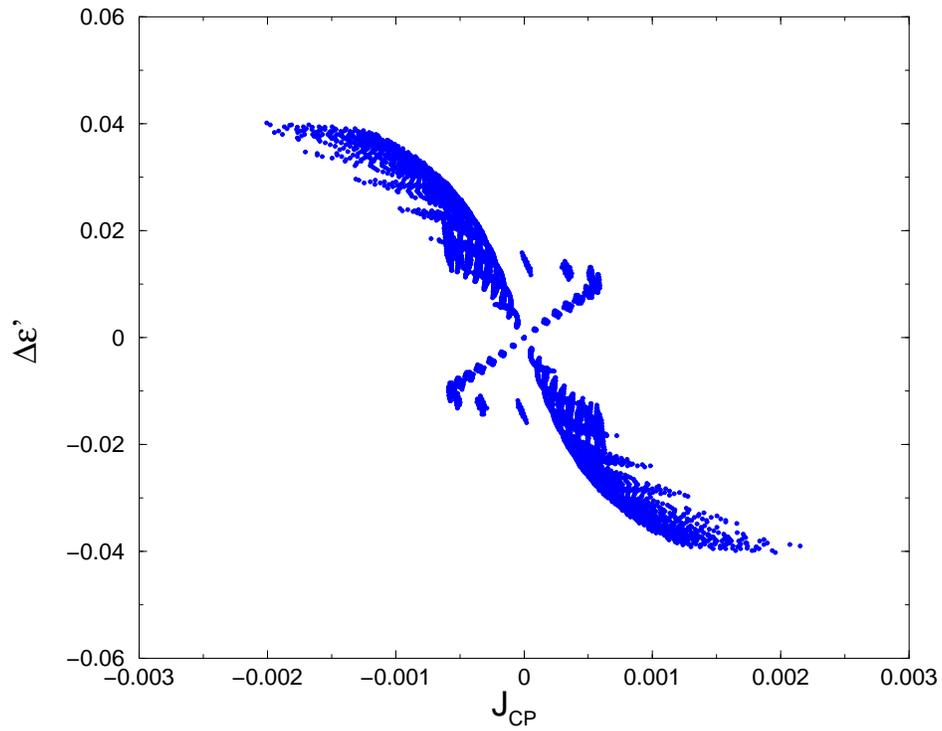}}
\caption{\label{fig-r1-Jcp-lpg} Correlation between 
the amount of leptogenesis 
and the leptonic Jarlskog invariant in Model I. 
}
\end{figure}

Once the parameters $(t,s,\alpha_{L},r)$ are determined, there are no more 
adjustable parameter. 
Using this data set, we can then predict the (13) element of the MNS 
matrix, $\left|U_{e3}\right|$, the leptonic Jarlskog 
invariant, $J_{\mbox{\tiny CP}}$, the matrix element for neutrinoless double beta decay, 
$\left< m_{ee} \right>$, and the amount of leptogenesis.  
Fig.~\ref{fig-r1-ue3-uatm} shows the correlation between 
the deviation of the atmospheric mixing angle from $\pi /4$ 
and the predicted value for $\left| U_{e3} \right |$, 
which is in the range of $(0.5-3)\times 10^{-3}$. 
The current experimental upper bound 
for $\left|U_{e3}\right|$ is $0.122$~\cite{Maltoni:2004ei}, and     
an improvement on this bound can be achieved in 
the very long baseline neutrino experiment~\cite{Diwan:2003bp}. 
Fig.~\ref{fig-r1-Jcp-uatm} shows 
the correlation between 
$1-\sin^{2}2\theta_{\mbox{\tiny atm}}$ and the predicted value for 
$J_{\mbox{\tiny CP}}$, which ranges from $0$ to $0.002$; in addition, 
a large value for $J_{\mbox{\tiny CP}}$ implies 
a large deviation for $1-\sin^{2}2\theta_{\mbox{\tiny atm}}$. 
Fig.~\ref{fig-r1-ue3-uatm} and Fig.~\ref{fig-r1-Jcp-uatm} 
also show that the 
deviation of the atmospheric mixing angle from $\pi /4$ 
is negligibly small, which is the main difference between this model 
and Model II. 
In Fig.~\ref{fig-r1-Jcp-nulb}, we show the correlation between the leptonic 
Jarlskog invariant and the prediction for neutrinoless double beta decay 
matrix element, which ranges between $5 \times 10^{-4}$ to 
$3 \times 10^{-2} \; \mbox{eV}$. Except for the region around 
$J_{\mbox{\tiny CP}} \simeq 0$, the value for $\left< m_{ee} \right>$ 
increases as the value of $J_{\mbox{\tiny CP}}$.  
The total amount of lepton number asymmetry, 
$\epsilon_{\mbox{\tiny total}} = \epsilon^{\Delta_{L}}$,  
is proprtional to $\Delta \epsilon^{\prime}$, defined as
\begin{equation}
\Delta \epsilon^{\prime} 
= \frac{3}{16\pi} \frac{f_{1}^{0}}{v_{L}}  \cdot  
 \frac{
 \mbox{Im} \biggl(  \mathcal{M}_{D}  \left(v_{L} f \, e^{i\alpha_{L}} 
\right)^{\ast} 
\mathcal{M}_{D}^{T} 
 \biggr)_{11} }{ ( \mathcal{M}_{D} \mathcal{M}_{D}^{\dagger} )_{11} } 
= \frac{\epsilon^{\Delta_{L}}}{\beta} \; .
\end{equation}
In deriving the above expression, we have used the property that 
the mass of the lightest RH neutrino, $M_{1}$, 
is proportional to $(f_{1}^{0}/v_{L})$, where $f_{1}^{0}$ is 
the smallest eigenvalue of the matrix $f$. 
So for fixed $\beta$, the ratio of $\Delta \epsilon^\prime$ to 
$\epsilon^{\Delta_{L}}$ is universal for all data points. Thus it suffices  
to consider $\Delta \epsilon^\prime$ when extracting the correlation.
In order for the lepton number asymmetry not to be washed out by the scattering 
processes, the out-of-equilibrium condition, characterized by the ratio of 
the decay rate of the lightest RH neutrino, $\Gamma_{1}$, 
to the Hubble constant at temperature equal to its mass, 
$ H |_{T=M_{1}} $, 
\begin{equation}
\gamma \equiv 
\frac{ \Gamma_{1} }{ H |_{T=M_{1}} } = 
\frac{ M_{\mbox{\tiny Pl} } }{ (1.7) (32\pi) \sqrt{g_{\ast}} v^{2}}
\cdot 
\frac{\left(\mathcal{M}_{D}\mathcal{M}_{D}^{\dagger}\right)_{11}}{M_{1}} 
< 1 \; ,
\end{equation}
with $\sqrt{g_{\ast}}$ being the number of relativistic degrees of freedom, 
must be satisfied. This condition can be re-written as 
\begin{equation} \label{ooequ}
\frac{\left(\mathcal{M}_{D}\mathcal{M}_{D}^{\dagger}\right)_{11}}
{M_{1}} < 0.01 \; eV \; .
\end{equation}
In our model, the quantity on the left hand side is given by
\begin{equation}\label{ratio}
\frac{\left(\mathcal{M}_{D}\mathcal{M}_{D}^{\dagger}\right)_{11}}
{M_{1}} \simeq 2 s \frac{\left( O_{R} \right)_{12}^{2}}{f_{1}^{0}} 
\left(\frac{m_{c}}{m_{t}}\right)^{2} v_{L} \; ,
\end{equation} 
which is highly suppressed by the factor $\left(m_{c}/m_{t}\right)^{2}$. 
In addition, as $f_{1}^{0} \sim t$ is of order $\mathcal{O}(0.1)$,   
which reflects the fact that the hierarchy in the Majorana mass matrices 
is small, it does not off-set the suppression from $\left(m_{c}/m_{t}\right)$. 
We have checked numerically and found that this quantity is of order 
$\mathcal{O}(10^{-7}) \; eV$ for all points. 
Thus the condition given in Eq.~(\ref{ooequ}) is satisfied 
and consequently there are no effects due to dilution.    
Fig.~\ref{fig-r1-Jcp-lpg} shows the correlation between the leptonic 
Jarlskog invariant and the amount of leptogenesis, 
charaterized by $\Delta \epsilon^\prime$. As both $\Delta \epsilon^\prime$ 
and $J_{\mbox{\tiny CP}}$ are proportional to $\sin\alpha_{L}$, the plot has a reflection 
symmetry in the second and the forth quadrants. This is not surprising as 
both $\Delta \epsilon^\prime$ and $J_{\mbox{\tiny CP}}$ are proportional 
to $\sin\alpha_{L}$. 
And similar to Fig.~\ref{fig-r1-Jcp-nulb}, a large value for $J_{\mbox{\tiny CP}}$ 
implies a large value for $\Delta \epsilon^\prime$,  
in the region when $\left|J_{\mbox{\tiny CP}}\right|>0.0005$.
To generate the observed amount of the baryon asymmetry of the Universe 
(BAU), 
$n_{b}/s_{e} \sim 10^{-10}$, where $n_{b}$ and $s_{e}$ are respectively 
the baryon number and entropy, requires the lepton number 
asymmetry $\epsilon_{\mbox{\tiny total}}$ 
to be of the order of $10^{-8}$. As the parameter $s=q^{2}/2\beta$ is of 
order $\mathcal{O}(10^{2} \sim 10^{3})$, if $q$ assumes a natural 
value $\sim 1$, $\beta$ is roughly $\mathcal{O}(10^{-2} \sim 10^{-3})$.
This leads to a total lepton number asymmetry $\epsilon_{\mbox{
\tiny total}} = \beta \Delta \epsilon^{\prime}$ of order $\mathcal{O}
(10^{-4} \sim 10^{-5})$, sufficient to generate the observed BAU. 
We note that an amount of lepton number asymmetry 
$\epsilon_{\mbox{\tiny total}} \sim 10^{-8}$ corresponds to 
a leptonic Jarlskog invariant $|J_{\mbox{\tiny CP}}| \sim 10^{-5}$.  
A value of $\beta$ in the range of $(10^{-2} \sim 10^{-3})$ corresponds 
to a $SU(2)_{R}$ breaking scale of $v_{R} \sim (10^{12}-10^{13})$ GeV. 
And, due to the fact that the hierarchy in the Majorana mass matrices 
is small, the mass of the lighest RH neutrino $M_{1}$ is lighter 
than $v_{R}$ only by about one order of magnitude. The scale of $v_{R}$ 
in our model is consistent with the bounds given  
in Ref.~\cite{Sahu:2004ny}.

\subsection{Model II}

Assuming the type-II see-saw term, $f_{ij}v_{L}$, has the following 
form, 
\begin{equation}\label{f}
f_{ij} = \left(
\begin{array}{ccc}
A & B & -B 
\\
B & D+\frac{A}{2} & D-\frac{A}{2}
\\
-B & D-\frac{A}{2} & D+\frac{A}{2}
\end{array} 
\right) \; .
\end{equation}
where 
\begin{equation}
A = \frac{1}{2} \left( f_{1}^{0} + f_{2}^{0} \right) \; , \quad
B = \frac{1}{ 2 \sqrt{2} } \left( f_{2}^{0} - f_{1}^{0} \right) \; , \quad
D = \frac{1}{2} f_{3}^{0} \; , 
\end{equation}
with $f_{1,2,3}^{0}$ being the eigenvalues of the matrix $f$. 
This matrix gives rise to exactly bi-maximal mixing pattern, and a vanishing 
value for $\left| U_{e3} \right|$. Adding the conventional type-I see-saw term,
\begin{equation}
M_{\nu}^{I} = -\frac{v_{L}}{\beta v^{2}} \left( M_{D}^{T}f^{-1} M_{D}\right)
\equiv - s v_{L} \frac{\left( P^{T} f^{-1} P \right) }{q^{2}} \; ,
\end{equation}
where $s = q^{2}/\beta$ in this case, gives rise to a 
deviation from maximal solar mixing, and a non-vanishing value 
for $\left| U_{e3} \right|$.
The $(33)$ element of $M_{D}$ dominates the Dirac neutrino mass matrix.  
Hence if we set $m_{u}=m_{c}=0$ in the matrix $P_{ij}$, 
the conventional type-I see-saw mass term contributes only to the $(33)$  
element in the total effective neutrino mass matrix; this contribution 
called $x$ is given by  
$x=\frac{s}{4} \; \biggl(
\frac{1}{f_{1}^{0}}+\frac{1}{f_{2}^{0}} + \frac{2}{f_{3}^{0}} \biggr)$. 
In this approximation, 
the Jarlskog invariant, $J_{\mbox{\tiny CP}}$, 
can be obtained analytically; it is given by, 
\begin{equation}
J_{\mbox{\tiny CP}} 
=  - \frac{1}{16} 
\frac{ \left(\Delta m_{21}^{0}\right)^{2}  \left(m_{1}^{0}-m_{2}^{0}\right)
\left( m_{2}^{0} - m_{3}^{0} \right) \left( m_{3}^{0} - m_{1}^{0} \right)}
{\Delta m_{21}^{2} \Delta m_{31}^{2} \Delta m_{32}^{2}} x \sin 
\left( \alpha_{L} \right) \; ,
\end{equation}
and the total amount of lepton number asymmetry is given by, 
\begin{equation}
\epsilon^{\Delta_{L}} = 
-\frac{3 \sqrt{2}}{16\pi} \frac{M_{1}}{v^{2}} 
\left( D+ \frac{A}{2} \right) \sin \alpha_{L}\; .   
\end{equation}
So both quantities to the leading order 
are proportional to $\sin\alpha_{L}$. 

\begin{figure}
{\center
\includegraphics[scale=0.445,angle=270]{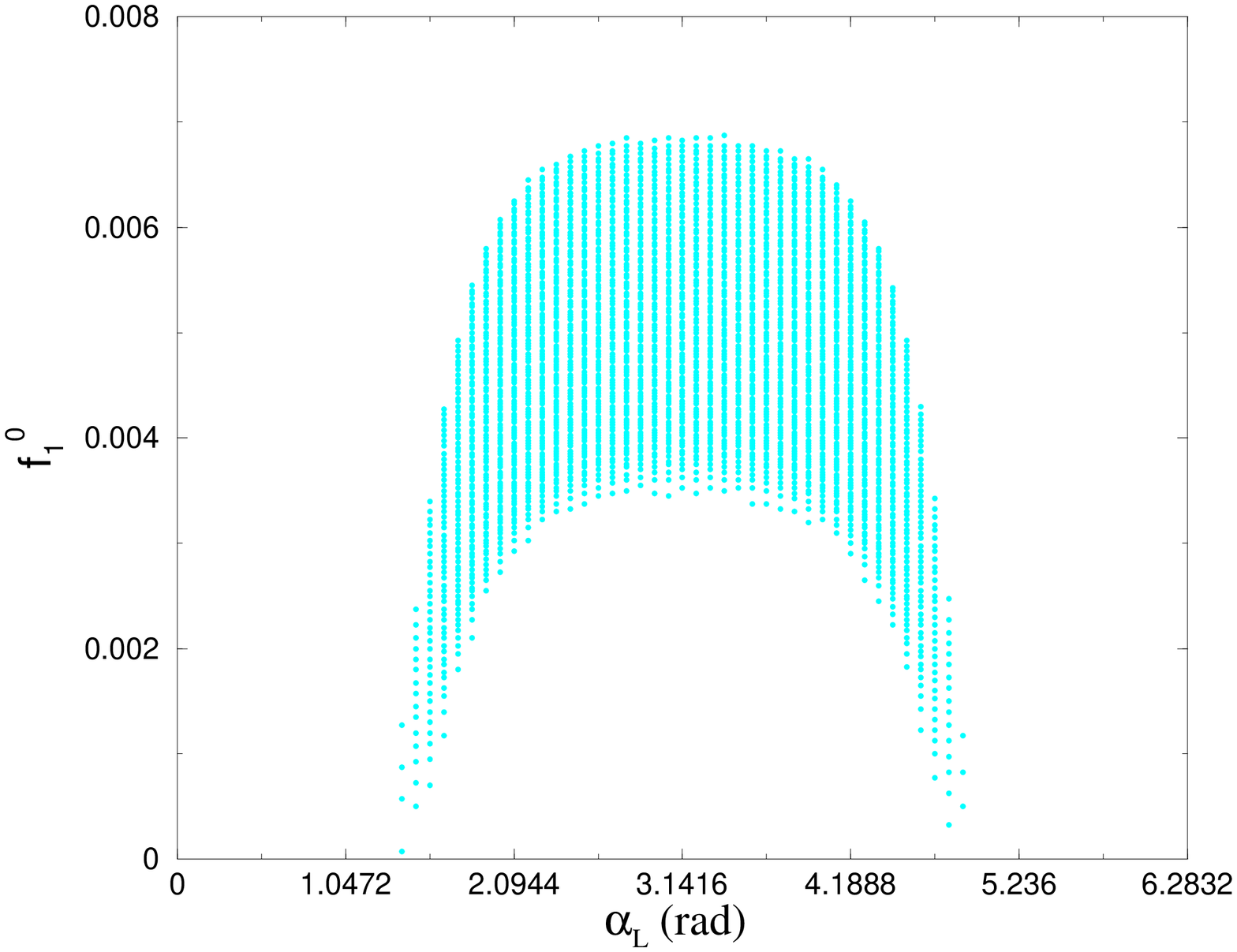}
\includegraphics[scale=0.445,angle=270]{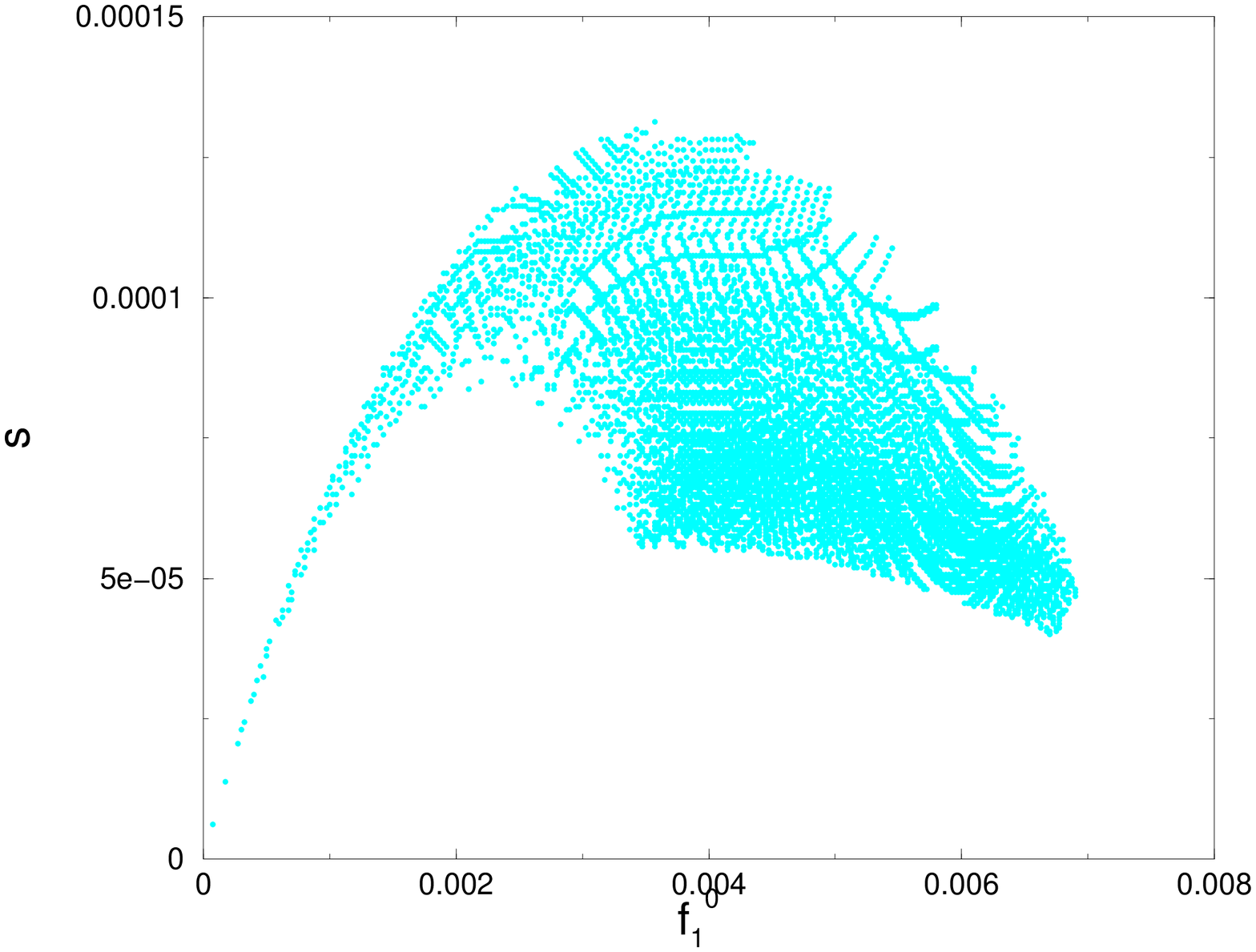}
\includegraphics[scale=0.445,angle=270]{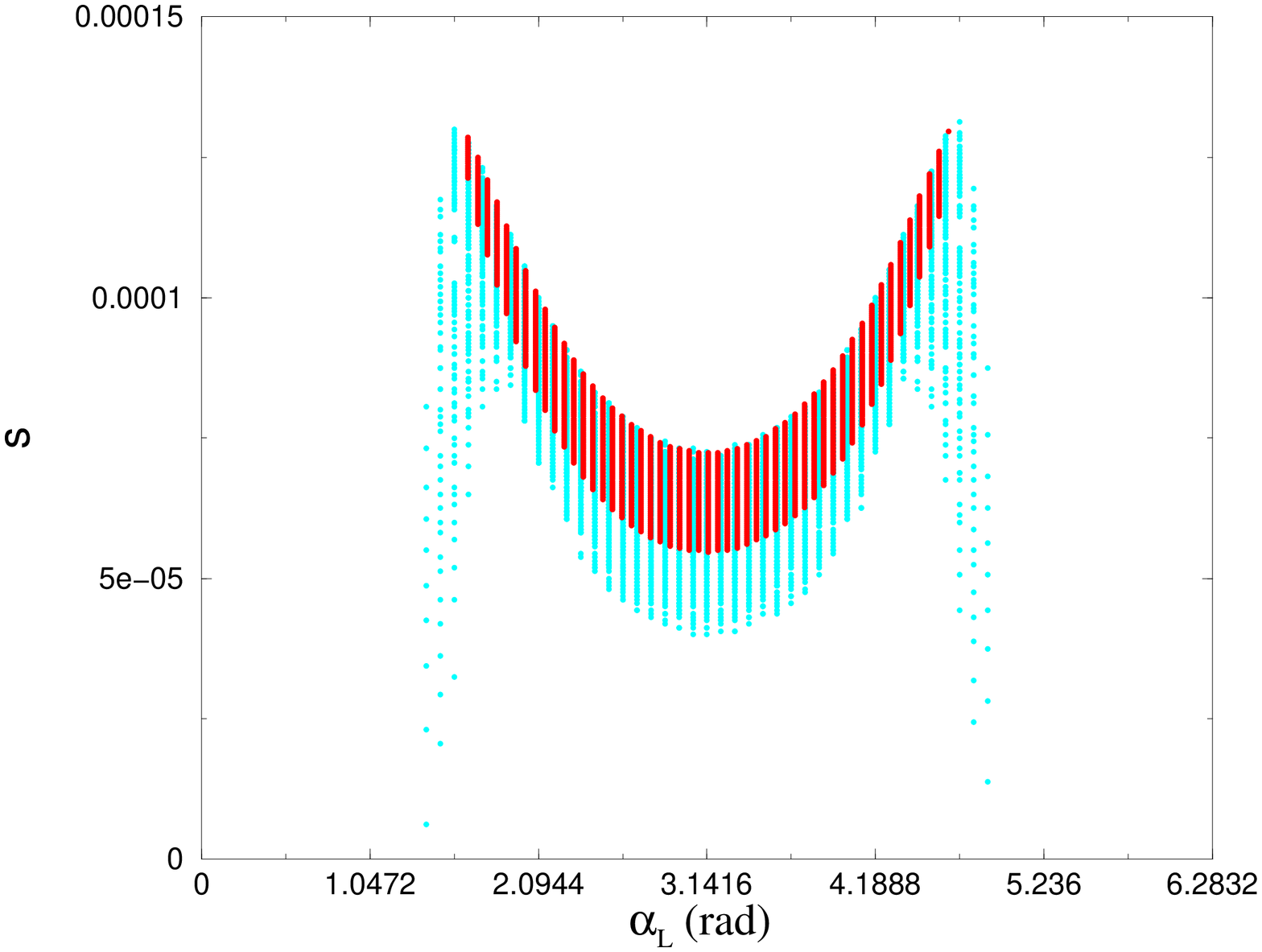}}
\caption{\label{fig-r2-a-s} Full allowed parameter space on 
the $(f_{1}^{0},\alpha_{L},s)$ plane in Model II. The shaded 
area in cyan (light shade) is the full allowed region, while the area 
in red (dark shade) corresponds to $f_{1}^{0} = 0.00424$. 
}
\end{figure}

\begin{figure}
{\center
\includegraphics[scale=0.55,angle=270]{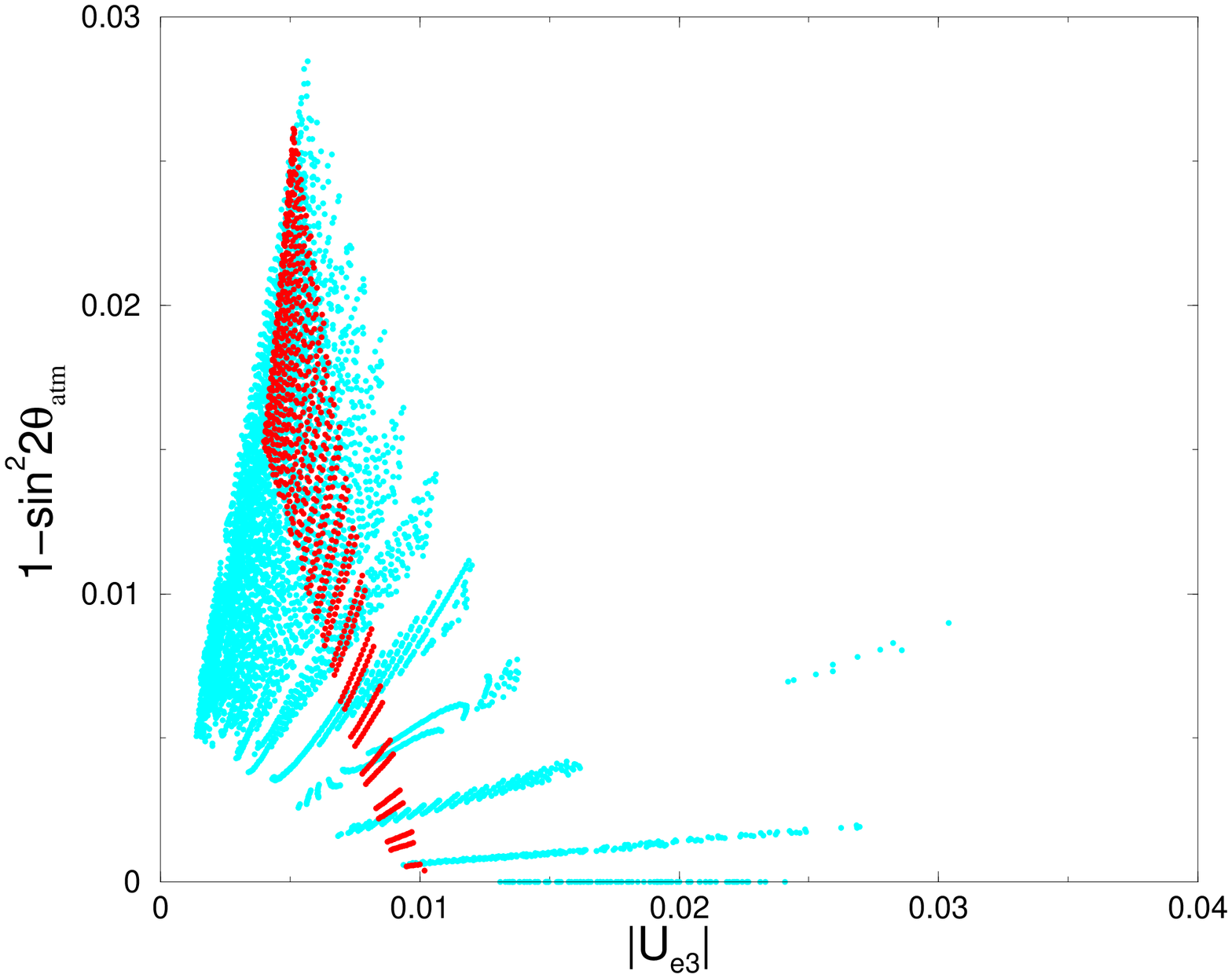}}
\caption{\label{fig-r2-ue3-uatm} Correlation between 
the matrix element of the leptonic mixing matrix, $\left| 
U_{e3} \right|$, and the deviation of the atmospheric mixing 
angle $\sin^{2}2\theta_{\mbox{\tiny atm}}$ from $1$ in Model II. 
The shaded area in cyan (light shade) is the full allowed region, 
while the area in red (dark shade) corresponds to 
$f_{1}^{0} = 0.00424$. 
}
\end{figure}

\begin{figure}
{\center
\includegraphics[scale=0.55,angle=270]{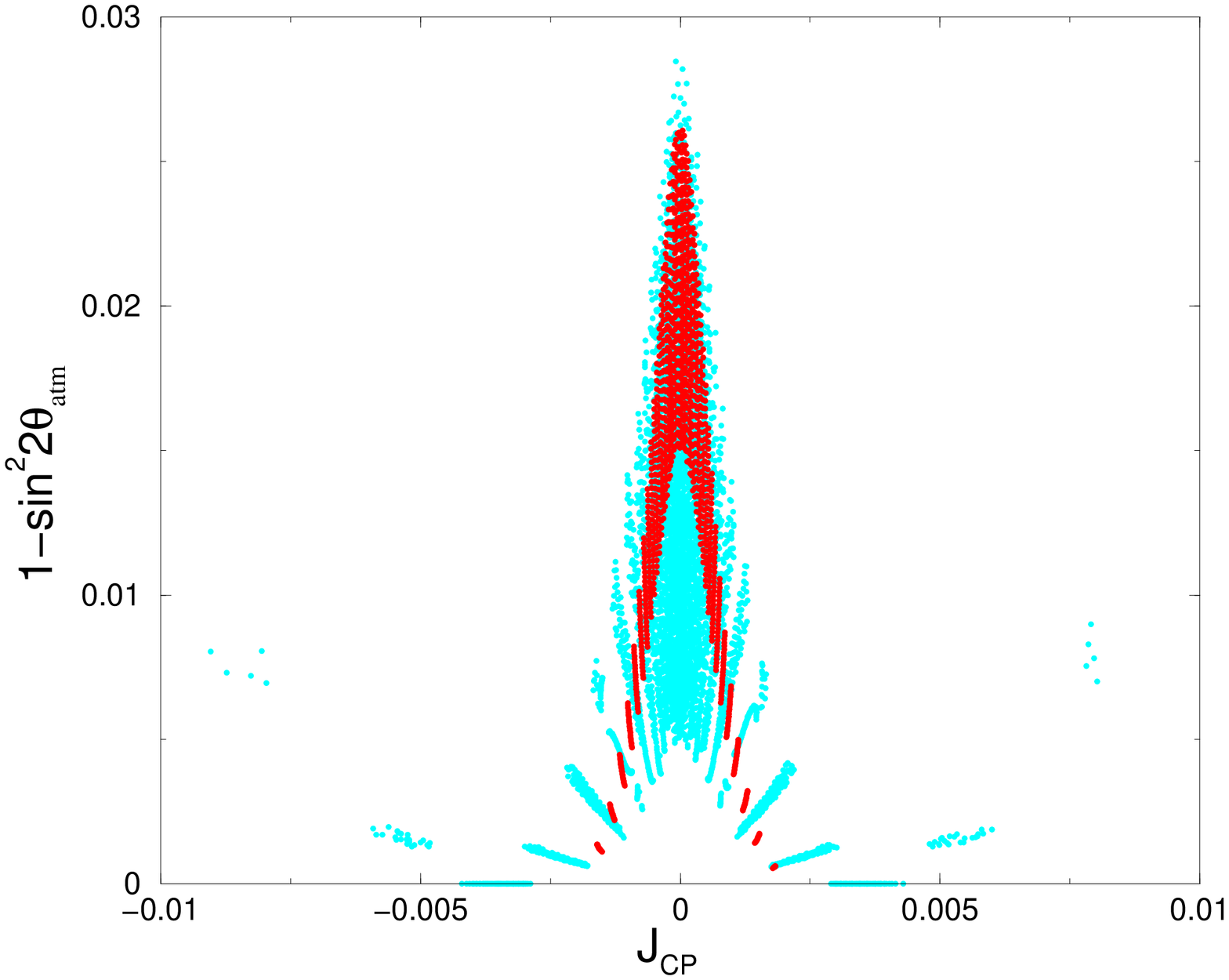}
\includegraphics[scale=0.55,angle=270]{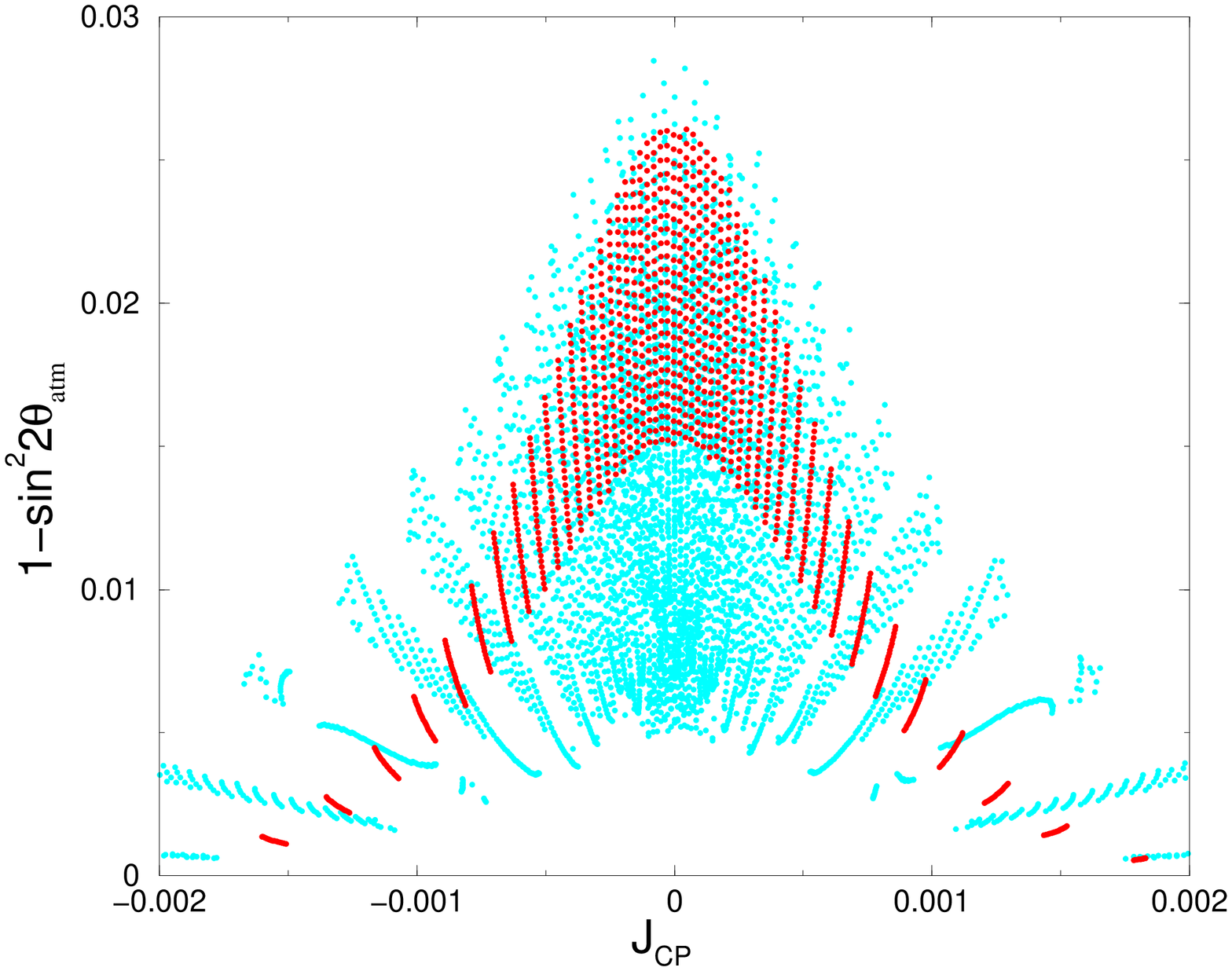}}
\caption{\label{fig-r2-Jcp-uatm} Correlation between 
the leptonic Jarlskog invariant and 
the deviation of the atmospheric mixing 
angle $\sin^{2}2\theta_{\mbox{\tiny atm}}$ from $1$ in Model II. 
The shaded area in cyan (light shade) is the full allowed region, 
while the area in red (dark shade) corresponds to 
$f_{1}^{0} = 0.00424$.}
\end{figure}

\begin{figure}
{\center
\includegraphics[scale=0.55,angle=270]{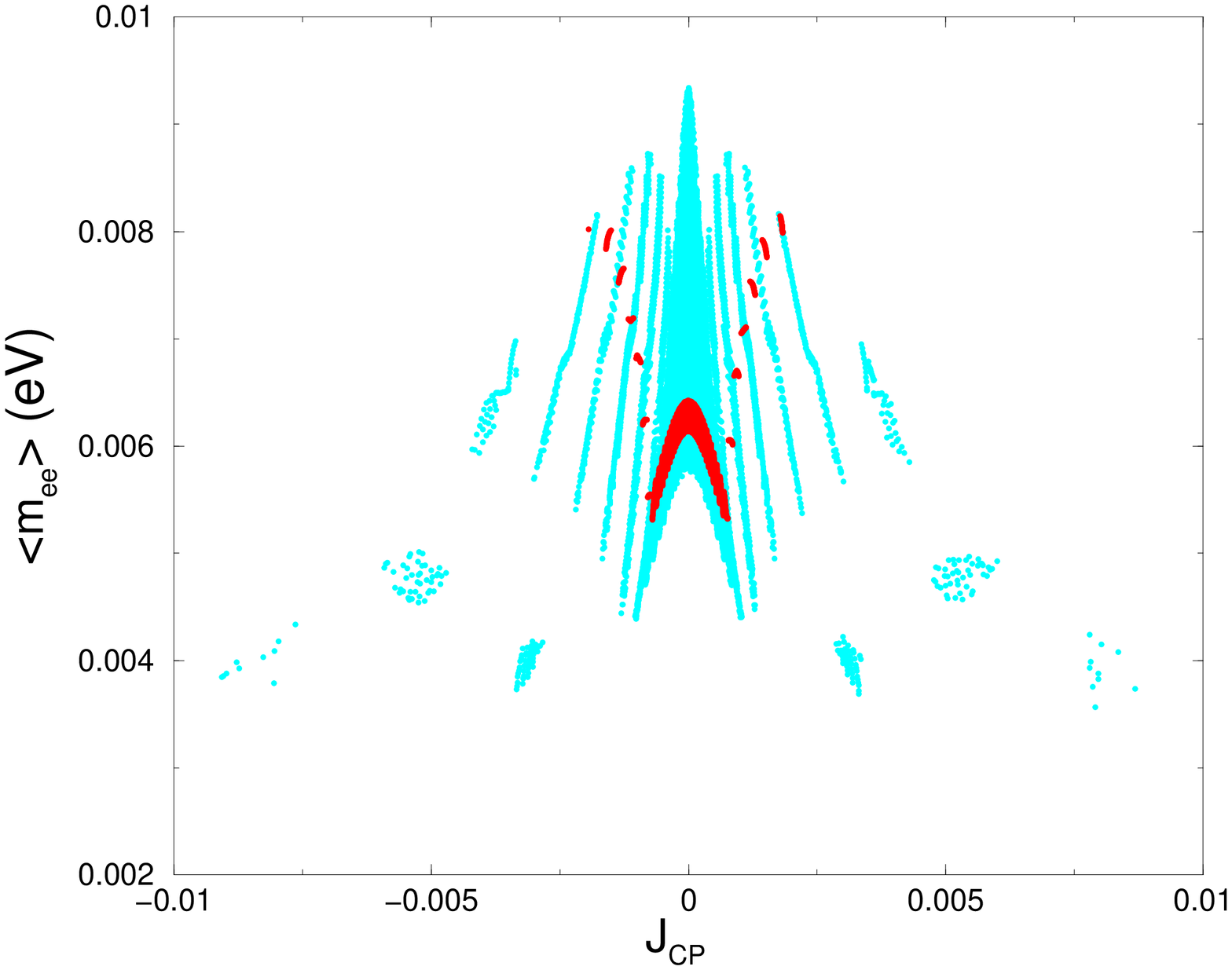}
\includegraphics[scale=0.55,angle=270]{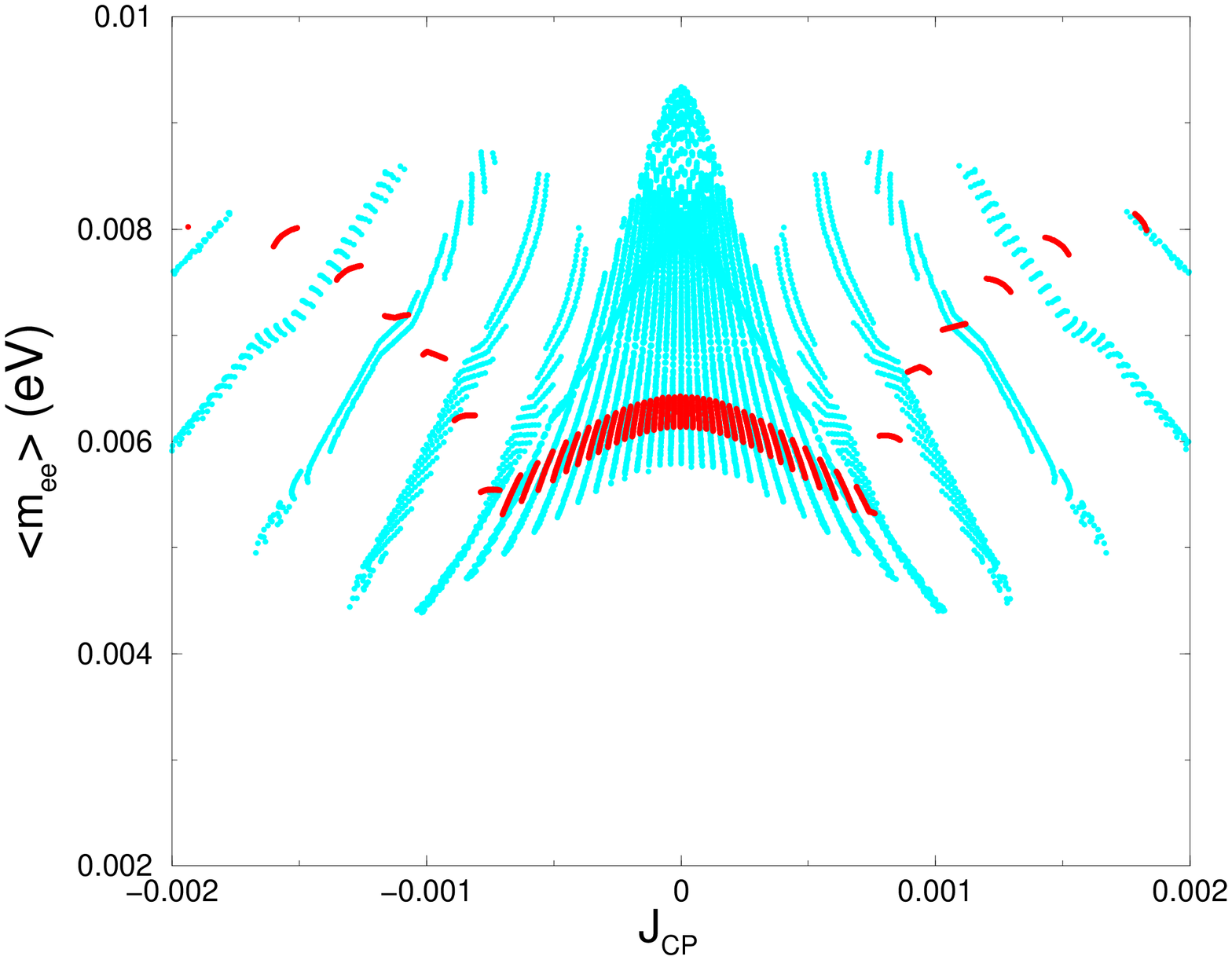}}
\caption{\label{fig-r2-Jcp-nulb} Correlation between 
the matrix element of neutrinoless double beta decay, $\left< 
m_{ee} \right>$, and the leptonic Jarlskog invariant in Model II. 
The shaded area in cyan (light shade) is the full allowed region, 
while the area in red (dark shade) corresponds to 
$f_{1}^{0} = 0.00424$.}
\end{figure}

\begin{figure}
{\center
\includegraphics[scale=0.55,angle=270]{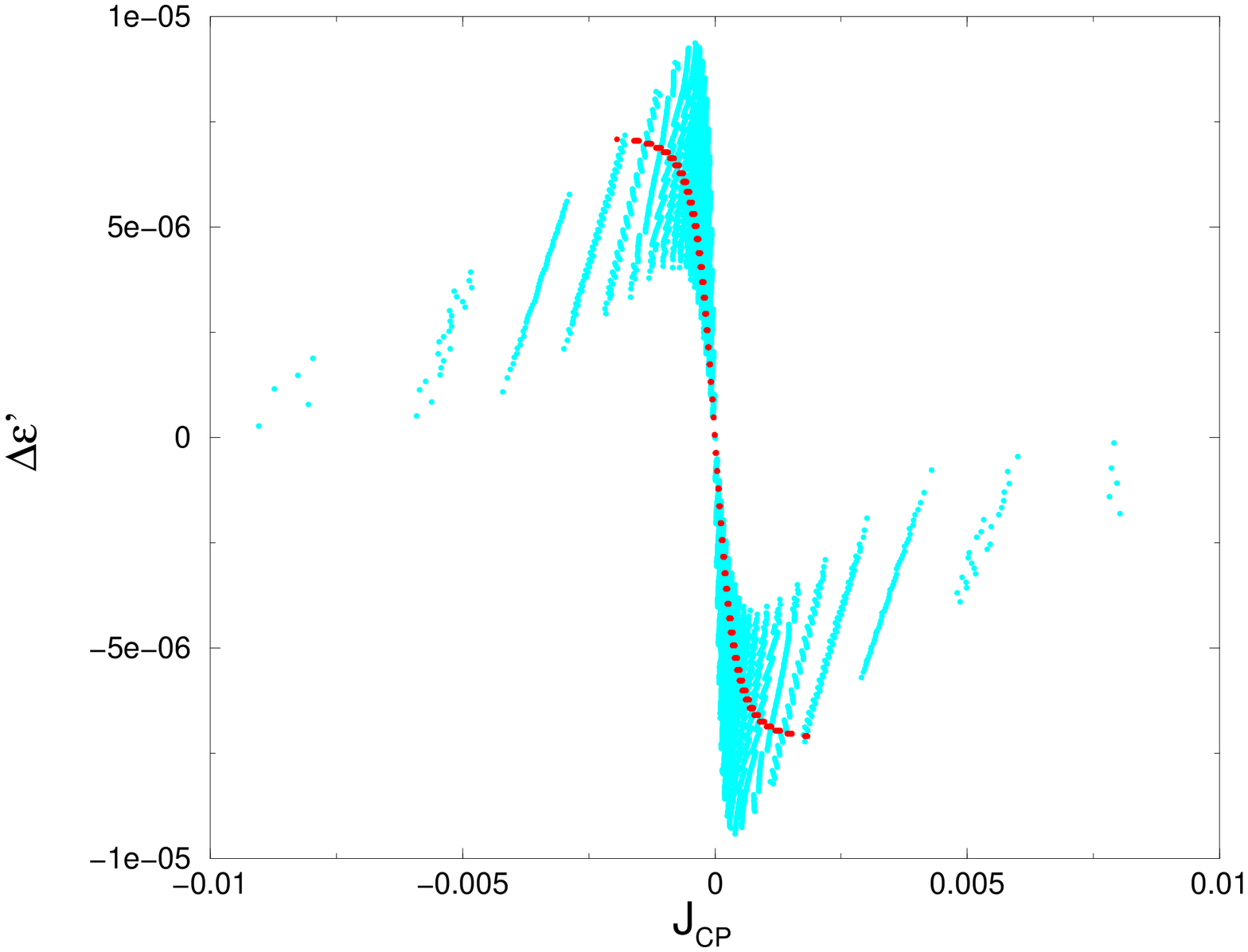}
\includegraphics[scale=0.55,angle=270]{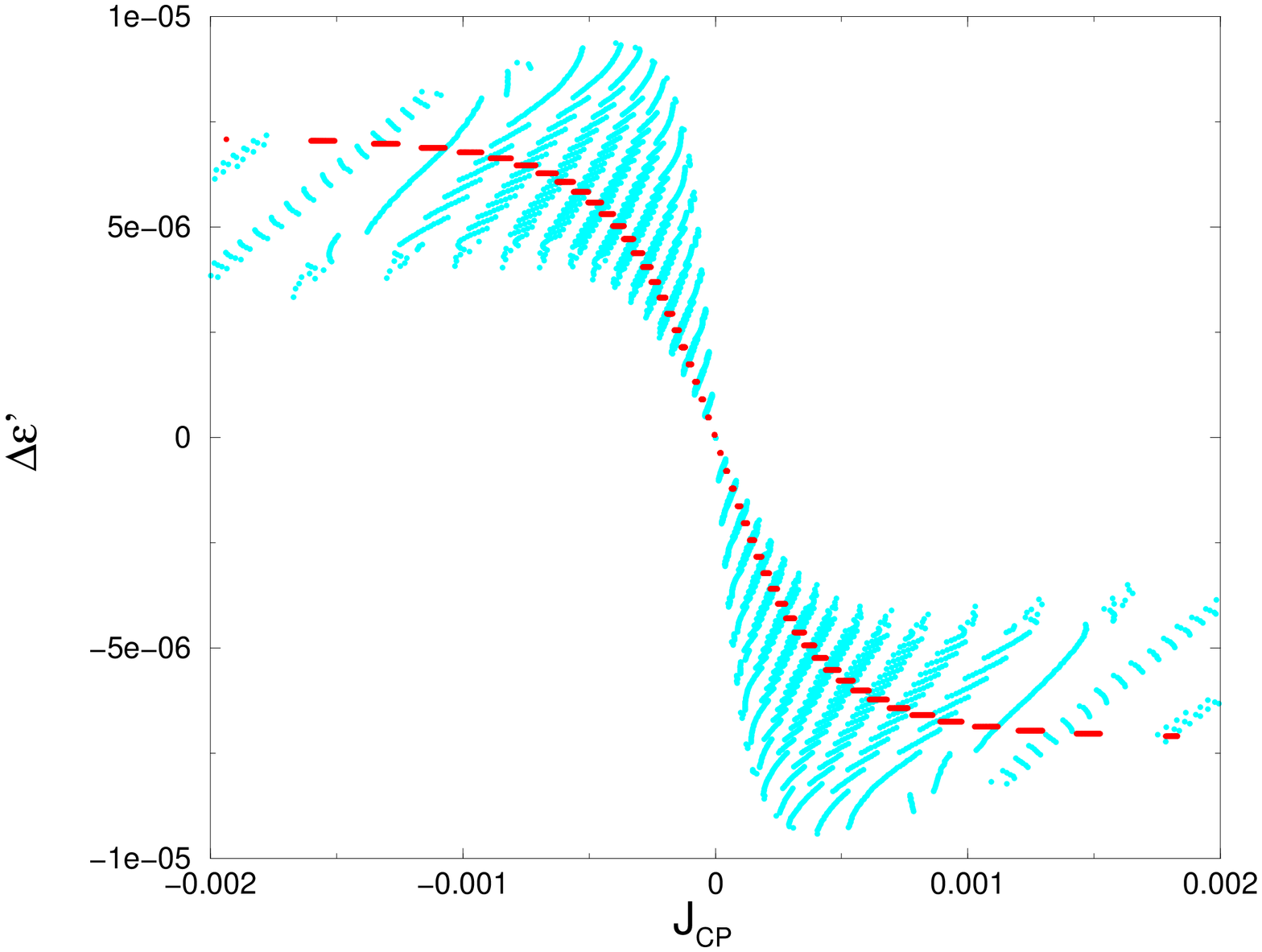}}
\caption{\label{fig-r2-Jcp-lpg} Correlation between 
the amount of leptogenesis 
and the leptonic Jarlskog invariant in Model II. 
The shaded area in cyan (light shade) is the full allowed region, 
while the area in red (dark shade) corresponds to 
$f_{1}^{0} = 0.00424$.}
\end{figure}

As we assume the hierarchical mass pattern for the neutrinos, the two heavier 
eigenvalues of the matrix $f$ can be approximated as   
$f_{3}^{0} v_{L} \sim \sqrt{\Delta 
m_{\mbox{\tiny atm}}^{2}}$ and $f_{2}^{0} v_{L} \sim \sqrt{\Delta 
m_{\odot}^{2}}$. 
Hence we choose $f_{2}^{0} = 0.01$ and $f_{3}^{0} = 0.049$ 
in our analysis with $v_{L} \sim 1\;$ eV, and treat $f_{1}^{0}$, 
along with $s$ and $\alpha_{L}$, as a free parameter. 
For the Dirac neutrino mass matrix $M_{D}$, 
we set $m_{u}/m_{t} = (0.22)^{8}$ and $m_{c}/m_{t} = (0.22)^{4}$.
Following the steps described in Model I, we search the 
full allowed region for $(f_{1},s,\alpha_{L})$, which is shown 
in Fig.~\ref{fig-r2-a-s}, and  
the corresponding $SU(2)_{L}$ triplet VEV $v_{L} = r \times  
(1 \; \mbox{eV})$ with $r = (0.847 - 1.21)$.  
The correlation between $\left| U_{e3} \right |$, $J_{CP}$ 
and $1-\sin^{2}2\theta_{\mbox{\tiny atm}}$, which, contrary to Model I, 
can be as large as $0.03$, are shown in 
Fig.~\ref{fig-r2-ue3-uatm} and Fig.~\ref{fig-r2-Jcp-uatm}, 
respectively. 
The predicted range of $\left| U_{e3} \right |$ 
is $\mathcal{O}(0.001-0.01)$ and that of 
$J_{\mbox{\tiny CP}}$ is $(0-0.03)$. Unlike in Model I,  
a small value for $\left|J_{\mbox{\tiny CP}}\right|$ implies 
a large deviation for $1-\sin^{2}2\theta_{\mbox{\tiny atm}}$.
In Fig.~\ref{fig-r2-Jcp-nulb}, we show the correlation between the leptonic 
Jarlskog invariant and the prediction for neutrinoless double beta decay 
matrix element, the range of which is $(3-10) \times 10^{-3}
\; \mbox{eV}$. Except for the region around 
$J_{\mbox{\tiny CP}} 
\simeq 0$, a large value of $\left|J_{\mbox{\tiny CP}}\right|$ implies a large 
value for $\left< m_{ee} \right>$.  
The correlation between the leptonic 
Jarlskog invariant and the amount of leptogenesis, 
which is proportional to $\Delta \epsilon^\prime$, 
is shown in Fig.~\ref{fig-r2-Jcp-lpg}, which shows 
a reflection symmetry in the second and the 
forth quadrants. A  large value for $\left|J_{\mbox{\tiny CP}}\right|$ 
implies a large value for $\Delta \epsilon^\prime$, 
which reaches a plateau when $\left|J_{\mbox{\tiny CP}}\right| > 0.0005$. 
A natural way to have $s$ of order $\mathcal{O}(10^{-4})$   
is to have $\beta \sim 1$ and $q \sim 10^{-2}$. 
With $\beta \sim 1$, the total amount of lepton number 
asymmetry $\epsilon_{\mbox{\tiny}}$ can be as large as $10^{-5}$. 
This is sufficient for generating the obaserved 
BAU. The quantity given in Eq.~(\ref{ratio}) is of order $\mathcal{O}
(10^{-7})$ in this case, and thus the out-of-equilibrium is satisfied.
The $v_{R}$ in this case is $10^{15}$ GeV, and the mass of the lighest 
RH neutrino is given by $M_{1} \sim 10^{12}$  GeV.

\section{Conclusion}\label{cond}

In this paper, we have shown that there 
exist correlations among the CP violation 
in leptogenesis, neutrino oscillation 
and neutrinoless double beta decay 
in the minimal left-right symmetric model with 
spontaneous CP violation in which 
there are only two intrinsic CP violating phases to account 
for all CP violation in both the quark and lepton sectors. 
We construct two realistic models and exhibit such correlations 
explicitly. 
Even though these two specific models have very different 
predictions for the neutrino oscillation parameters, both models 
exhibit the feature that a large leptonic Jarlskog invariant 
implies a large value for leptogenesis. When $\left|J_{\mbox{\tiny CP}}\right| 
= 0$, leptogenesis vanishes. This statement is true for any model  
having non-vanishing $\left| U_{e3} \right|$.
Except for $\left|J_{\mbox{\tiny CP}}\right| \simeq 0$, in both models, 
a large value for $\left|J_{\mbox{\tiny CP}}\right|$ also implies 
a large value for the matrix element 
of the neutrinoless double beta decay.  
The connection between the CP violation in the leptonic 
sector and that in the quark 
sector is rather weak due to the large hierarchy in the 
bi-doublet VEV required by a realistic quark sector.

\begin{acknowledgments}
M-CC and KTM are supported, in part, 
by the U.S. Department of Energy under Grant No. DE-AC02-98CH10886 and 
DE-FG03-95ER40892, respectively. M-CC would also like to acknowledge 
Aspen Center for Physics, where part of this work was done, for its 
hospitality and for providing a very stimulating atmosphere. 
\end{acknowledgments}

\bibliography{lrmodel}

\end{document}